\documentclass[a4paper,11pt]{article}
\usepackage{jcappub} 
\usepackage{lineno}
\usepackage{multirow}
\usepackage{amsmath}
\usepackage{bm}
\usepackage[flushleft]{threeparttable}
\usepackage{booktabs} 
\usepackage{float} 
\usepackage[numbers]{natbib}
\usepackage{cancel}
\usepackage{caption}
\usepackage{subcaption}
\usepackage{graphicx}
\usepackage{hyperref} 
\usepackage{comment}

\makeatletter
\newcommand{\mref}[1]{
  \hyperref[#1]{\ensuremath{m_{\expandafter\@gobble\string#1}}}
}
\makeatother

\newcommand{{\pasj}}{Publications of the Astronomical Society of Japan}

\newcommand{\HI}{H\,\textsc{I}}

\author[a]{Vishrut Pandya,}
\author[a]{Leon Noble,}
\author[a]{Suman Majumdar,}
\author[b]{Debanjan Sarkar,}
\author[c,d]{Mohd Kamran,}
\author[a]{Abhirup Datta}

\emailAdd{vpandya081@gmail.com}

\affiliation[a]{Department of Astronomy, Astrophysics \& Space Engineering, Indian Institute of Technology Indore, Indore 453552, India}
\affiliation[b]{Department of Physics and Trottier Space Institute, McGill University, QC H3A 2T8, Canada}
\affiliation[c]{INAF–Astronomical Observatory of Trieste, Via G. B. Tiepolo 11, 34143 Trieste, Italy}
\affiliation[d]{IFPU–Institute for Fundamental Physics of the Universe, Via Beirut 2, 34151 Trieste, Italy}

\abstract{The redshifted 21-cm signal from neutral hydrogen (\HI) in the intergalactic medium (IGM) is a powerful probe of the Epoch of Reionization (EoR). Owing to the complex growth and morphology of ionized regions, the 21-cm brightness-temperature field becomes strongly non-Gaussian during the EoR, limiting the information captured by the standard power spectrum alone. While higher-order statistics such as the bispectrum can recover part of this information, they are computationally expensive and often less straightforward to interpret. In this work, we investigate marked statistics as an alternative framework for characterizing the EoR 21-cm signal. Using semi-numerical 21-cm simulations, we introduce a set of EoR-tailored mark functions, inspired by and extending existing marked-statistics ideas, and study both the power spectrum of the mark and that of the marked field. We show that suitably chosen marks can selectively enhance contributions from different IGM environments, capture additional non-Gaussian information beyond the standard power spectrum, and improve the statistical constraining power on EoR model parameters in a Fisher-matrix analysis. These results demonstrate that Fourier-space marked statistics, including both the power spectrum of the mark and that of the marked field, provide a computationally simple and flexible extension of standard two-point statistics for extracting astrophysical information from the EoR 21-cm signal.}

\begin{document}
\title{\boldmath Probing Reionization up to the Mark: \\
The Marked Power Spectrum to unveil the HI 21-cm signal from the EoR}

\maketitle
\flushbottom

\section{Introduction}

The Cosmic Dawn (CD) marks the emergence of the first stars and galaxies, whose radiation began to heat and ionize the neutral hydrogen (\HI) in the intergalactic medium (IGM). The subsequent transition of the Universe from a largely neutral state to an almost fully ionized one is referred to as the Epoch of Reionization (EoR). Our present understanding of this period has been shaped primarily by indirect observational probes, including measurements of the CMB Thomson-scattering optical depth \cite{Komatsu_2011, Plank_2016, Plank_2020}, absorption signatures in the spectra of high-redshift quasars \cite{Becker_2001, Fan_2003, Barnett_2017}, and the luminosity and clustering properties of ${\rm Ly}\alpha$ emitters \cite{Ouchi_2010, Zheng_2017, Taylor_2021}. While these observations have provided important constraints on the timing and global progression of reionization, they do not directly reveal key aspects of the process, such as the nature of the first ionizing sources or the morphology and growth of ionized regions. This motivates direct observational probes of the CD-EoR. Since the IGM during this era is rich in neutral hydrogen, the redshifted 21-cm signal arising from the hyperfine spin-flip transition of \HI\ offers a direct tracer of the state of the IGM and the structure of ionized regions \cite{Furlanetto_2006, Pritchard_2012}. Several ongoing experiments, including uGMRT \cite{Paciga_2013}, HERA \cite{Abdurashidova_20221}, LOFAR \cite{Mertens_2025}, and MWA \cite{Barry_2019}, together with the forthcoming SKAO \cite{Koopmans_2015}, have the potential to detect this signal from the EoR directly. Although no direct detection has yet been achieved, increasingly stringent upper limits on the spherically averaged 21-cm power spectrum have already been reported by these interferometric experiments \cite{Paciga_2013, Abdurashidova_20221, Mertens_2025, Barry_2019, Gehlot_2019, Mertens_2020, Abdurashidova_2022, acharya2024revisedlofarupperlimits, ceccotti2025upperlimits21cmsignal}. Recent work has also presented upper limits on the 21-cm bispectrum from MWA observations \cite{gill2025eor21cmbispectrumz82, Gill:2026svk}. In parallel, a variety of approaches have been explored to extract information beyond conventional low-order statistics \cite{Simpson_2011, gruen2015weaklensinggalaxytroughs, Paillas_2021, paillas2023cosmologicalconstraintsdensitysplitclustering, Banerjee_2020, Cheng__2020, eickenberg2022waveletmomentscosmologicalparameter, Cheng__2025, Rubira_2021, 2ptcollaboration2024parametermaskedmockdatachallenge}.


The redshifted \HI\ 21-cm differential brightness temperature during the EoR is expected to be strongly non-Gaussian \cite{Bharadwaj_2005, Iliev_2006, Mellema_2006}. Since the power spectrum is a complete statistical descriptor only for a Gaussian random field, it cannot by itself capture all of the information encoded in such a signal. Accessing this non-Gaussian information generally requires statistics beyond the two-point level, for example the bispectrum \cite{Majumdar_2018,Majumdar_2020, Watkinson_2022,Tiwari_2022, Noble:2024uzl, Mahida:2025teg}. While such higher-order statistics can provide important additional insight, their estimation and interpretation are often more challenging, both computationally and conceptually. This motivates the search for alternative summary statistics that can capture part of the non-Gaussian information in the EoR 21-cm signal while retaining the relative simplicity of two-point estimators. In this work, we investigate the \textit{marked power spectrum}, the Fourier-space counterpart of the \textit{marked correlation function}, as a framework for characterizing non-Gaussianity in EoR \HI\ 21-cm intensity maps.

A mark is, in general, a function of a local property of the field and acts as a transformation and/or reweighting of the field itself. Owing to its non-linear construction, the marked two-point statistic can encode information from higher-order correlations in addition to that contained in the original two-point function \cite{Ebina_2025}. The concept of the marked correlation function was introduced in \cite{https://doi.org/10.1002/mana.19841160115} and has since been applied in a wide range of contexts, from environment-dependent matter clustering to weak-lensing analyses. Marked statistics provide a useful framework for studying the connection between galaxy properties and environment. In \cite{sheth2005markedcorrelationsgalaxyformation, Beisbart_2000, beisbart2002markcorrelationsrelatingphysical,10.1111/j.1365-2966.2006.10196.x, Skibba_2012}, quantities such as luminosity and star-formation rate have been used as marks to study their relation to galaxy clustering. In \cite{beisbart2002markcorrelationsrelatingphysical, Gottl_ber_2002, Sheth_2005, White_2009}, marked correlation functions have been employed to investigate halo-model statistics and to help break degeneracies among halo-model parameters. Marked correlation functions have also been used to distinguish between modified-gravity scenarios \cite{PhysRevLett.107.071303, Winther_2012, PhysRevLett.114.251101, 10.1093/mnras/stx865, 10.1093/mnras/sty1335, 10.1093/mnras/stz009, Hern_ndez_Aguayo_2019, Armijo_2024, White_2016, aviles2021testingmodifiedgravitytheories}. They have further been used to quantify the Alcock-Paczynski effect, to assess the impact of redshift-space distortions, and to improve cosmological parameter constraints \cite{Yang_2020}. More recently, \cite{marinucci2024constrainingpowermarkedpower} presented an analytical study of the marked power spectrum in the context of primordial non-Gaussianity of the non-local type, while \cite{Ebina_2025} clarified its connection to higher-order correlation functions. The marked power spectrum has also been used to tighten constraints on neutrino masses \cite{PhysRevLett.126.011301}. In the context of galaxy surveys, it has been applied to clustering analyses and cosmological parameter inference in \cite{Massara_2023, massara2024scsimbigcosmologicalconstraints}. Marked power spectra have also recently been applied to the neutral-hydrogen field in the post-reionization regime \cite{kamran2024remarkable21cmpowerspectrum}. The first application of marked angular power spectra to HSC-Y1 weak-lensing data was presented in \cite{cowell2025constraintsmarkedangularpower}. Recent work such as \cite{ebina2026markedpowerspectrumpractical} has further demonstrated that suitably defined marked power spectra can help break parameter degeneracies by isolating higher-order information, reducing overlap with the standard power spectrum through a redefinition of the statistic, and exhibiting a smooth dependence on cosmology that is amenable to interpolation-based inference.


The non-Gaussian nature of the EoR \HI\ 21-cm signal evolves in time, reflecting both the growth of matter fluctuations and the inhomogeneous ionizing radiation field in the IGM. Together, these processes generate a range of environments with distinct spatial and statistical characteristics. In this work, we extend marked statistics to the EoR \HI\ 21-cm brightness-temperature field and, to our knowledge, present their first application in this context. By introducing suitable marks, we selectively modulate the contribution of different environments and use the marked power spectrum to quantify this environment-dependent clustering across spatial scales. A practical advantage of this approach is that it can be incorporated into an existing power-spectrum pipeline with only minimal modification: once a mark is defined, one can construct the transformed or reweighted field and estimate its two-point statistics in Fourier space. Previous studies have shown that certain non-linear transformations can partially Gaussianize the density field \cite{Neyrinck_2009, Neyrinck_2011}. In a related sense, the application of a mark is also conceptually connected to screening or clipping transformations of the density field \cite{Lombriser_2015,Simpson_2011}. Such transformations can therefore transfer part of the higher-order information of the field into suitably defined two-point statistics. In this way, marked statistics provide a promising route to improving constraints on EoR model parameters beyond those obtainable from the standard power spectrum alone.


The paper is organized as follows. In Section 2, we introduce the 21-cm marked-statistics framework, including the generation of the \HI\ differential brightness-temperature maps, the definition of the marks, and the marked summary statistics used in this work. In Section 3, we present our results, including the marked power spectra obtained using different transformations and the corresponding Fisher-matrix analysis of their information content relative to the standard power spectrum. We also compare the complementary information extracted from different environments and assess its impact on the forecasted constraints on the EoR model parameters. In Section 4, we discuss the main advantages, challenges, and limitations of the 21-cm-marked-power-spectrum approach, and summarize our conclusions. The appendices present the evolution of the 21-cm marked power spectrum and a discussion of additional mark functions considered in this work.

\section{The 21-cm Marked Statistics} 

\subsection{Simulating HI 21-cm differential brightness temperature}

We use the semi-numerical simulation code \texttt{ReionYuga}\footnote{\href{https://github.com/rajeshmondal18/ReionYuga}{https://github.com/rajeshmondal18/ReionYuga}} \cite{Choudhury_2009, Majumdar_2014, Mondal_2016}, which is based on the excursion-set formalism \cite{Furlanetto_2004}, to generate the redshifted 21-cm brightness-temperature maps analyzed in this work. We first run a Particle-Mesh (PM) ${\rm N}$-body simulation to generate the dark-matter density field at redshift $z = 7.52$. The simulation volume is $[215.04\,\mathrm{cMpc}]^3$, discretized with $3072^3$ grids and $1536^3$ particles. This corresponds to a dark-matter particle mass resolution of $\approx 10^8\,\mathrm{M}_\odot$ and a spatial resolution of $0.07\,\mathrm{Mpc}$. From the resulting dark-matter particle distribution, we identify collapsed halos using a Friends-of-Friends (FoF) algorithm with a linking length equal to $0.2$ times the mean inter-particle separation. Halos containing at least $10$ dark-matter particles are retained \cite{Majumdar_2014}, corresponding to a minimum halo mass of $1.09 \times 10^9\,\mathrm{M}_\odot$. We then generate the 21-cm brightness-temperature maps using \texttt{ReionYuga}. The model is characterized by three EoR parameters: the ionizing efficiency, ${\rm N}_{\rm ion}$, which sets the proportionality between the number of ionizing photons and the dark-matter halo mass; the minimum halo mass, ${\rm M}_{\rm h,min}$, which defines the lower halo-mass threshold for sources participating in reionization; and the mean free path of ionizing photons, ${\rm R}_{\rm mfp}$. The parameter values are chosen such that the resulting maps correspond to a mass-averaged neutral fraction of $\sim 0.5$ at $z = 7.52$.

\subsection{Marking the signal using a suitable mark}
\begin{figure}[htbp]
    \centering
    \includegraphics[width=\linewidth]{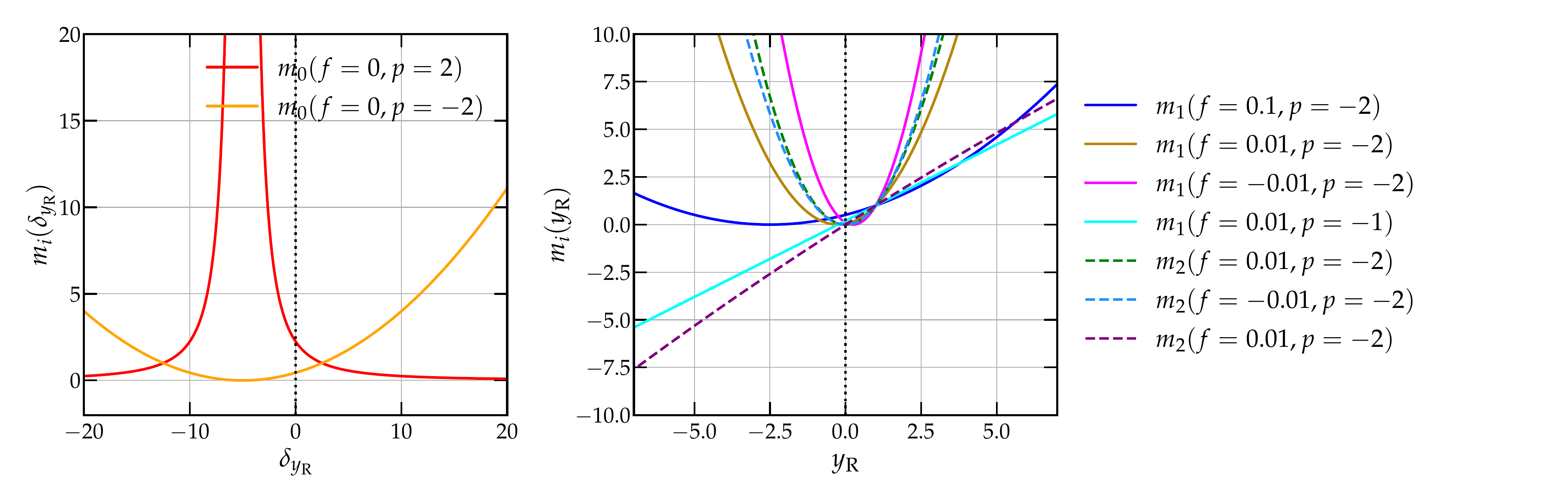}
    \caption{Illustration of the idea of mark, as transformation of a form of input field $y_{\rm R}$ and $\delta_{y_{\rm R}}$ for a set of mark parameters. Different colors correspond to different transformations, highlighting different ways to encode information from environments in the map. The left plot demonstrates the simple, versatile nature of mark \mref{m0} for selecting contributions from complementary environments (ionized and neutral regions) in the IGM. The right plot shows the ability of marks \mref{m1} and \mref{m2} to 'boost' contributions from environments proportionately ($p=\pm1$) and irrespective of their nature ($p=\pm2$).}
    \label{fig:mark_functional_forms}
\end{figure}

The central idea of marked statistics is to apply a suitable \textit{mark}, i.e. a transformation of the input field that depends on one of its local properties, and to estimate summary statistics of the transformed field, such as the correlation function or the power spectrum, in order to characterize the target features of the original field. This framework has been widely used both in real space, through the \textit{marked correlation function}, and in Fourier space, through the \textit{marked power spectrum}, primarily in the context of density-weighted three-dimensional matter or galaxy fields. In analogy with such density-weighted constructions, we extend the idea of marking to EoR 21-cm brightness-temperature maps, defining the mark as a function of the 21-cm brightness-temperature fluctuation. The choice of the functional form of the mark is not unique and depends on the scientific objective. In the present work, we use the mark both to enhance the contribution from particular environments in the IGM and more generally as a transformation designed to improve information extraction from EoR 21-cm maps. Motivated by \cite{White_2016}, we introduce a simple mark that can be used to isolate the contributions from ionized and neutral regions in the 21-cm signal:
\begin{equation} \label{m0}
    m_0(\mathbf{x},z;\text{R},f,p) = \Bigg[ \frac{f + 1}{f + 2 + \delta_{\text{T}_{\text{b}}}(\mathbf{x},z,\text{R})} \Bigg]^p,
\end{equation}
where $\delta_{\text{T}_{\text{b}}}(\mathbf{x}, z, \text{R})$ is the brightness-temperature contrast $ \delta_{{\rm T}_{\rm b}}(\mathbf{x, z}) = \delta{\rm T}_{\rm b}(\mathbf{x}, z) / \langle \delta {\rm T}_{\rm b}(z) \rangle -  1$ smoothed with a spherical top-hat filter of radius ${\rm R}$. The free parameters ${\rm R}$, $f$, and $p$ define the \textit{mark parameters}. Their roles may be interpreted as follows. The smoothing scale ${\rm R}$ sets the characteristic scale below which fluctuations are suppressed. The thresholding parameter $f$ controls the strength of the transformation, while the exponent $p$ determines which environments are preferentially emphasized. As illustrated in the left panel of Figure \ref{fig:mark_functional_forms}, positive values of the exponent (e.g. $p=2$) enhance the contribution from ionized regions, whereas negative values (e.g. $p=-2$) enhance that from neutral regions. Therefore, the same functional form, \mref{m0}, can be used with different parameter choices to selectively tune the contribution from different environments in the map.

Similarly, we propose two other marks, as more general transformations, that can systematically tune the contributions from both types of local environment in the map. This allows capturing the correlation of a point with its surroundings, and such a construction can be used to encode higher-order information that otherwise may not be captured by the standard power spectrum. With this motivation, we propose the following forms.  

\begin{equation}\label{m1}
    m_1(\mathbf{x},z; \text{R},f,p) =
    \Bigg[ \frac{1 + f \, \langle \delta \text{T}_{\text{b}}(z) \rangle}
    { f\,\langle \delta \text{T}_{\text{b}}(z) \rangle + 
    \Big[ \text{dT}_{\text{b}}(\mathbf{x},z)\Big]_\text{R} }\Bigg]^p ,
\end{equation}
 
 and 

\begin{equation}\label{m2}
    m_2(\mathbf{x},z; \text{R},f,p) =
    \Bigg[ \frac{1 + f\, \Big[ \text{dT}_{\text{b}}(\mathbf{x},z) \Big]}
    { f\, \Big[ \text{dT}_{\text{b}}(\mathbf{x},z) \Big] + 
    \Big[ \text{dT}_{\text{b}}(\mathbf{x},z)\Big]_\text{R} }\Bigg]^p
\end{equation}

where $\delta\text{T}_\text{b}(\mathbf{x},z)$ is the differential brightness temperature and $\text{dT}_\text{b}(\mathbf{x},z) = \delta\text{T}_\text{b}(\mathbf{x},z) - \langle \delta\text{T}_\text{b}(z)\rangle$ is the mean-subtracted brightness temperature. In interferometric observations, the mean (monopole) component of the signal is not captured, as interferometers do not measure the zero-baseline visibility \cite{McKinley_2018}. Functional forms, like \mref{m2}, can be used to analyze such data-sets, as \mref{m2} is a function of the mean-subtracted map itself. 

The mark parameters, in these marks, can be interpreted in the same way as the \mref{m0} where the thresholding parameter $f$ is weighted by global (\mref{m1}) or local (\mref{m2}) brightness temperature. The right panel in Figure \ref{fig:mark_functional_forms}, shows the response of marks \mref{m1} and \mref{m2} to the map. For $p=\pm1$, \mref{m1} and \mref{m2} proportionately increase and decrease the contributions from positive (neutral regions) and negative (ionized regions) temperature contrast respectively and that for $p=\pm2$, both contrasts are emphasized. This in turn increases the contribution of the local environment of a point in the map. As shown in \cite{Cowell_2024}, marks, as transformations, exhibit symmetries: apparently different marks can encode the same information content of the field. We exploit this property to restrict our analysis to a representative set of functional forms and mark parameters.

\begin{figure}[htbp]
    \centering
    
    \subcaptionbox{Mean subtracted map $({\rm dT}_{\rm b})$.\label{fig:dTm}}[0.45\linewidth]{
        \includegraphics[width=\linewidth]{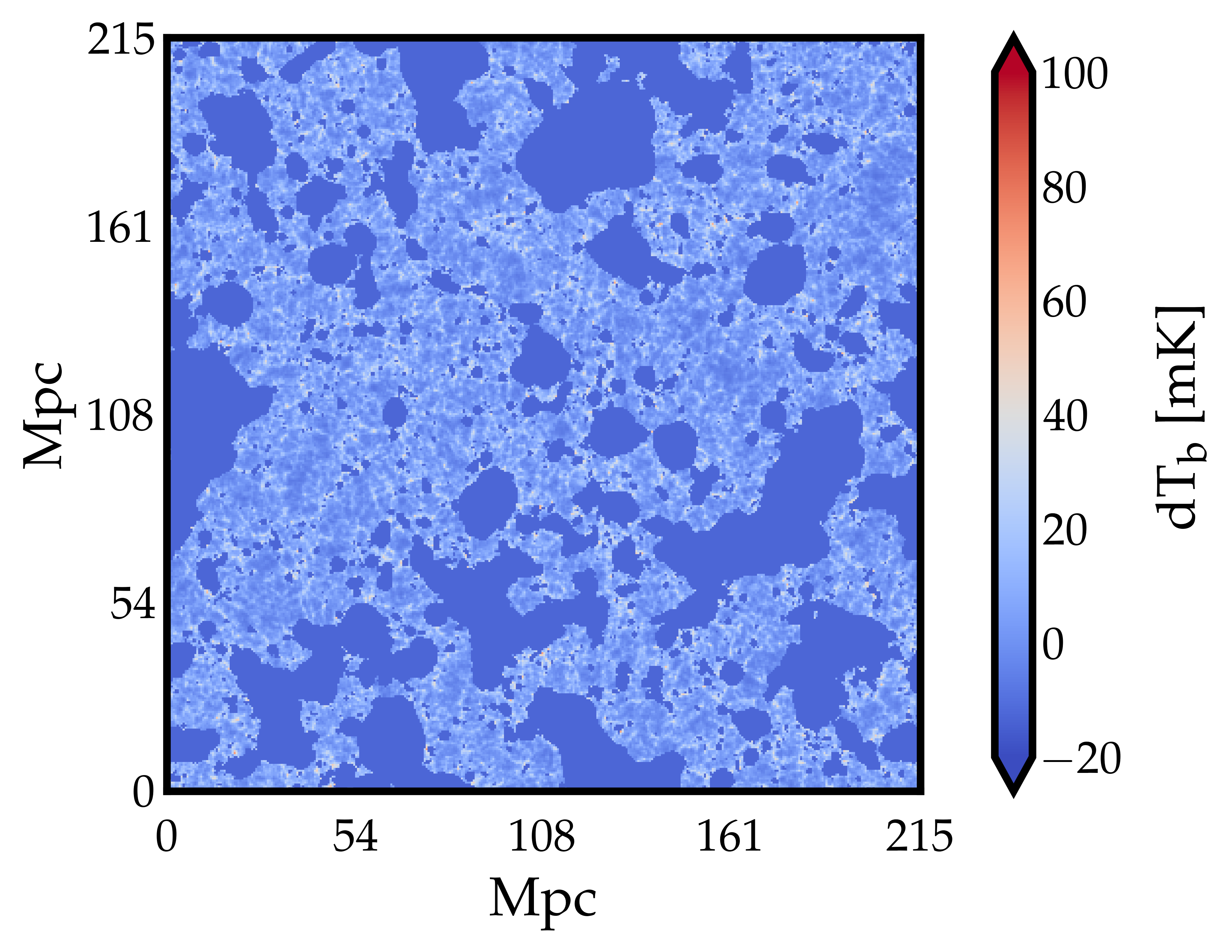}}
    \hfill
    \subcaptionbox{Brightness temperature contrast $(\delta_{{\rm T}_{\rm b}})$.\label{fig:delta_T}}[0.45\linewidth]{
        \includegraphics[width=\linewidth]{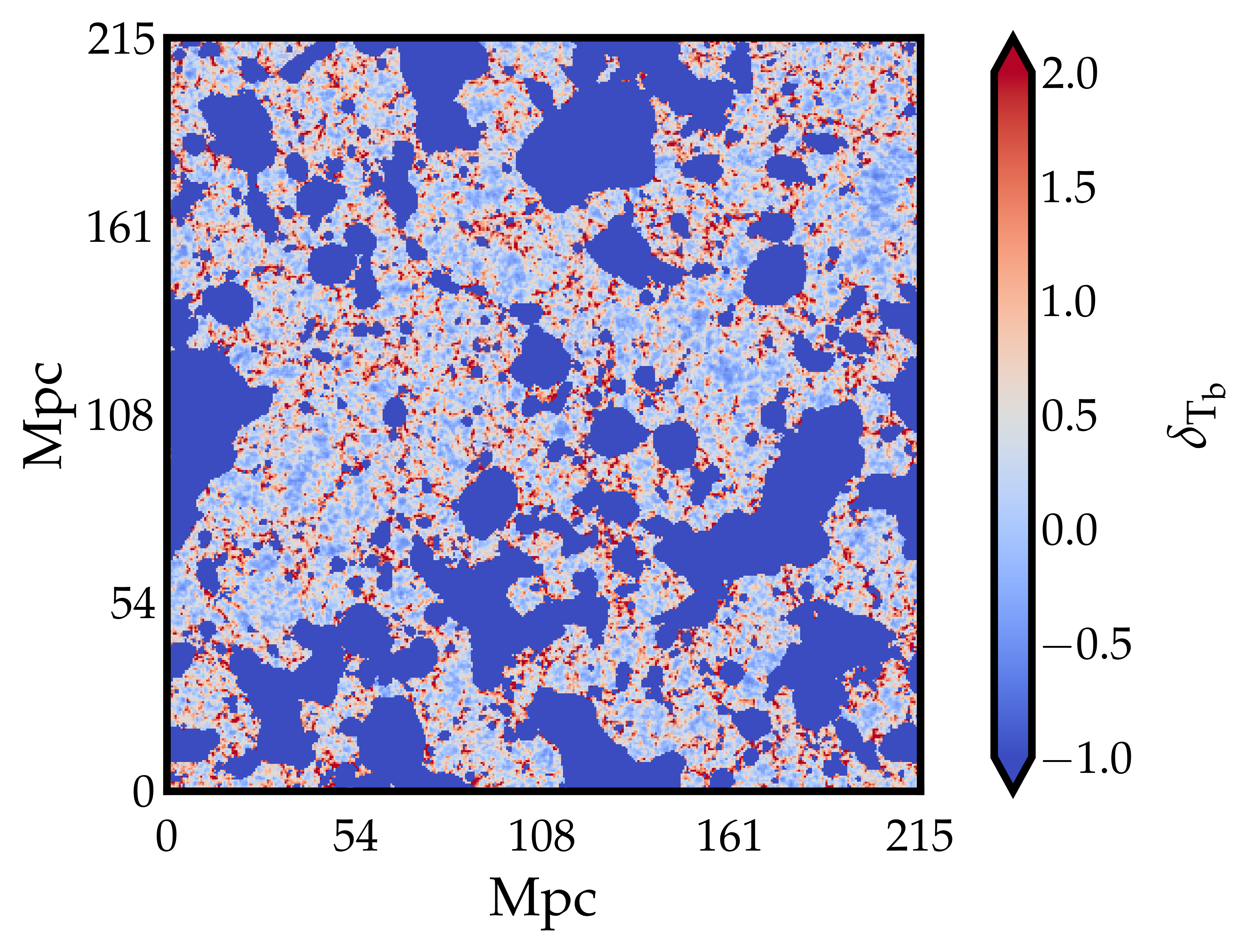}}
    
    \vspace{0.1em}
    
    \subcaptionbox{$m_0(\text{R}=1.68 \, \text{Mpc},f=0, p=2)$.\label{fig:marks_visualization_p2}}[0.45\linewidth]{
        \includegraphics[width=\linewidth]{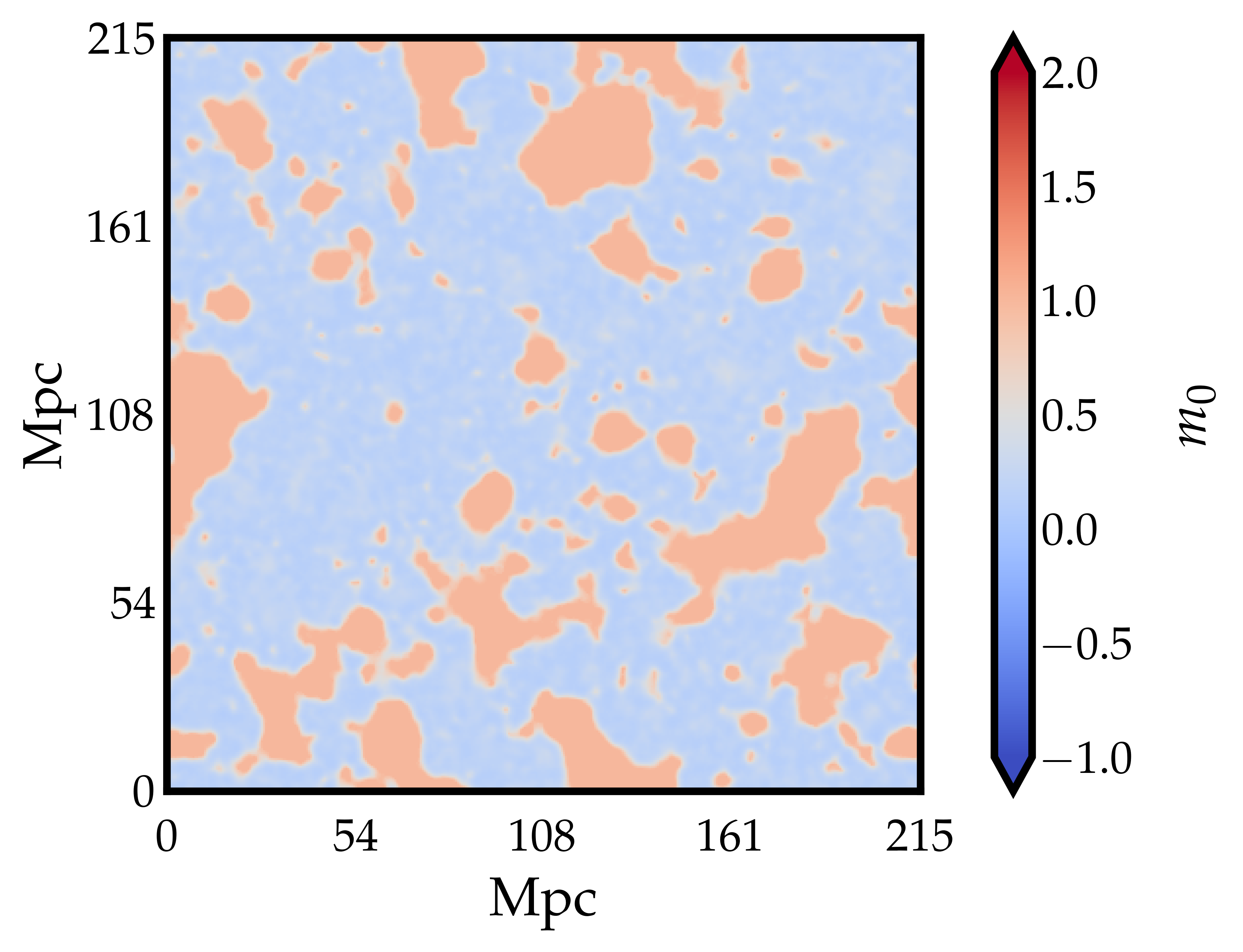}}
    \hfill
    \subcaptionbox{$m_0(\text{R}=1.68 \, \text{Mpc},f=0, p=-2)$.\label{fig:marks_visualization_pm2}}[0.45\linewidth]{
        \includegraphics[width=\linewidth]{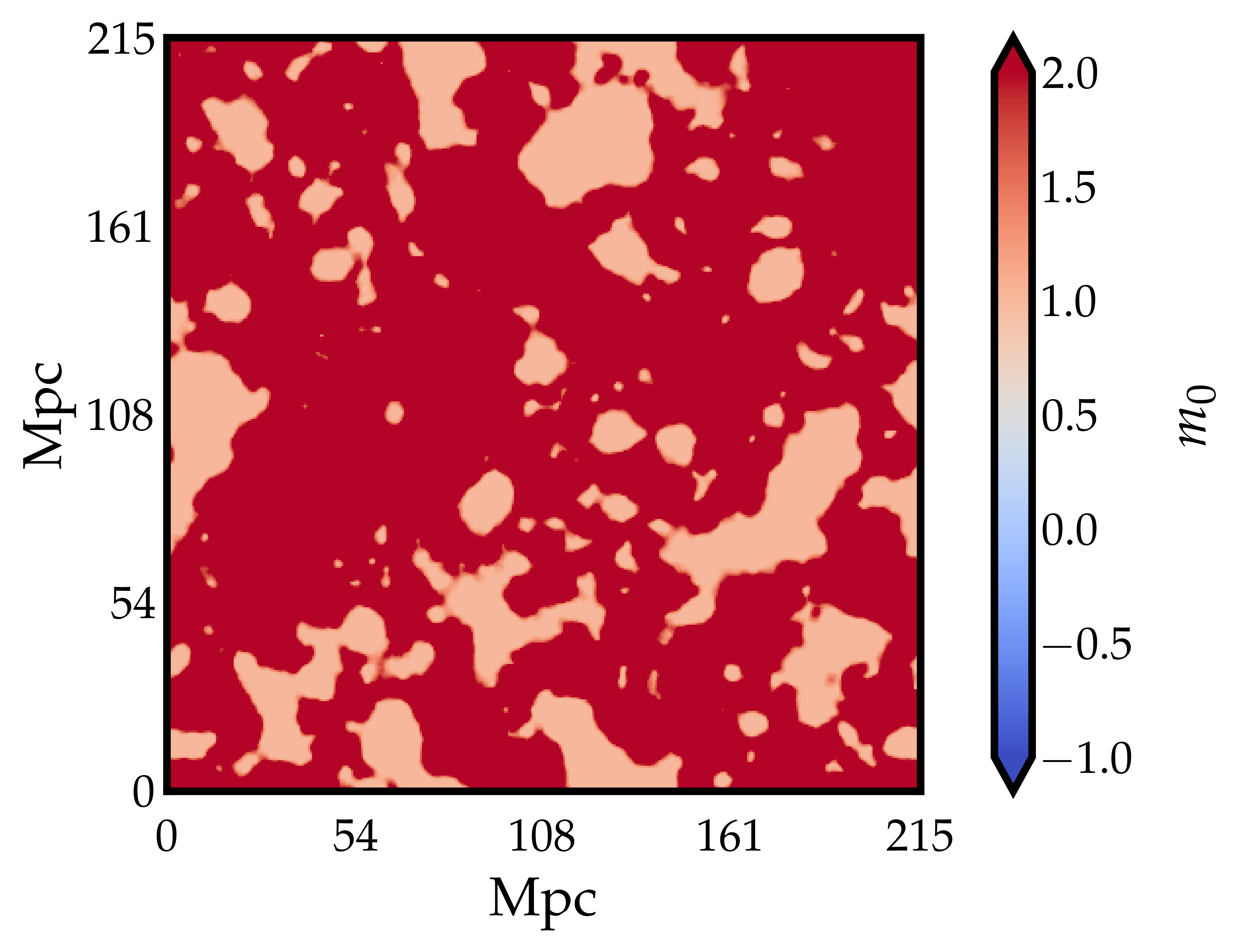}}
    
    \vspace{0.1em}
    
    \subcaptionbox{$m_1(\text{R}=1.68 \, \text{Mpc},f=0.01, p=-2)$.\label{fig:marks_visualization_m3}}[0.45\linewidth]{
        \includegraphics[width=\linewidth]{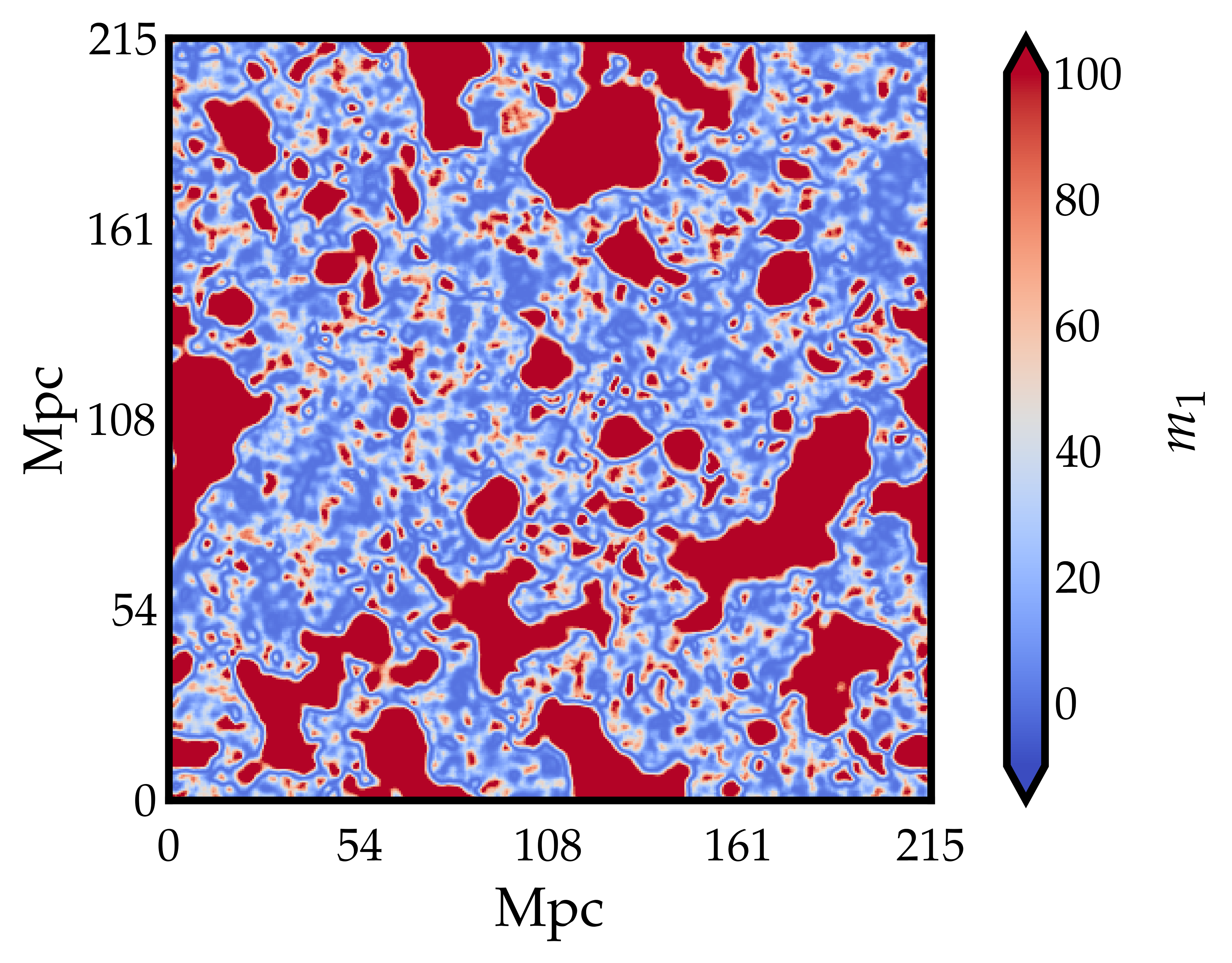}}
    \hfill
    \subcaptionbox{$\mathbf{M_1}(\text{R}=1.68 \, \text{Mpc},f=0.01, p=-2)$.\label{fig:marks_visualization_M3}}[0.45\linewidth]{
        \includegraphics[width=\linewidth]{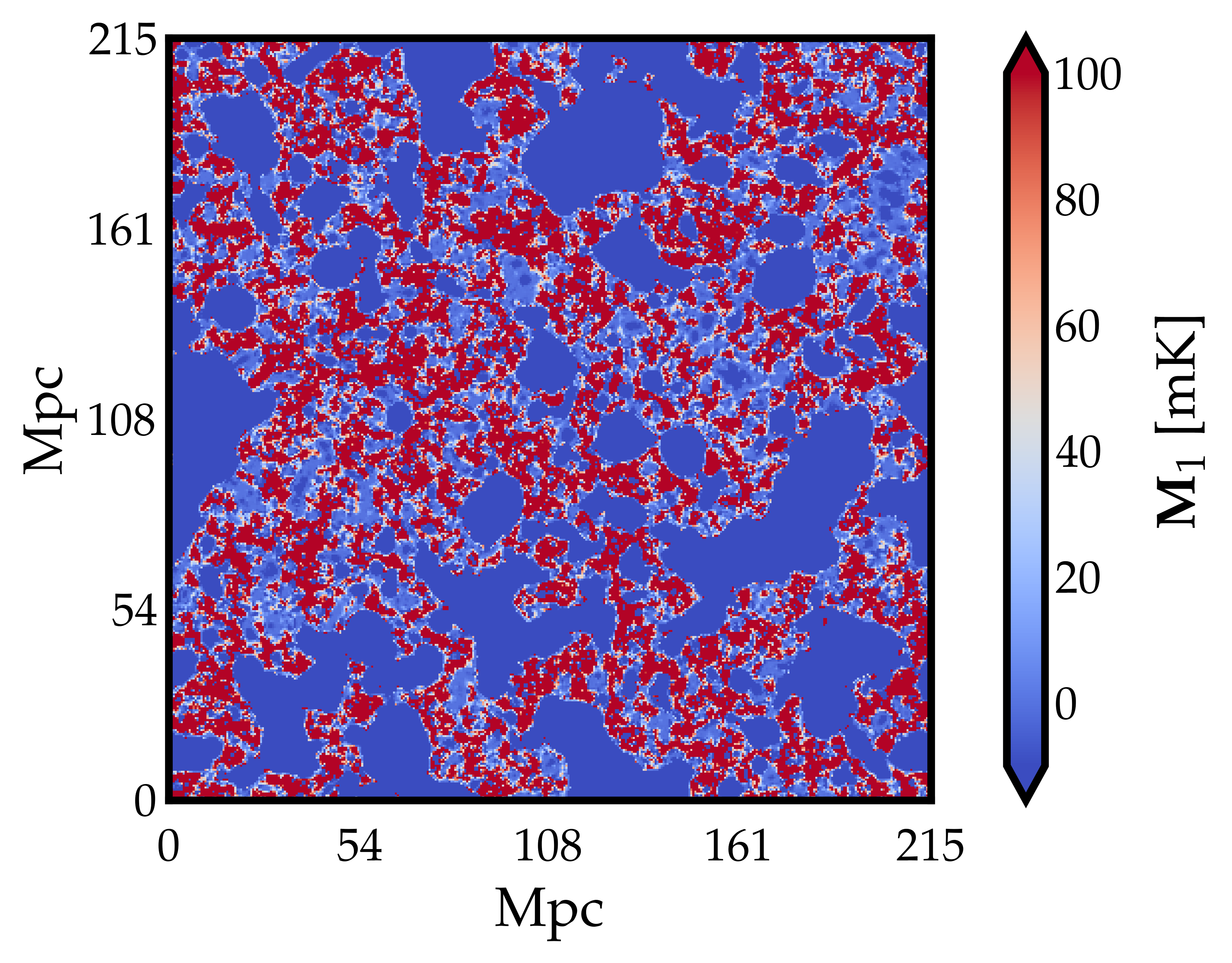}}

    \caption{Visual comparison of differential brightness temperature maps and some versions of marks at $z=7.52$ (mass averaged neutral fraction $\bar{x}_{\text{HI}} = 0.5$). The top panels show the mean-subtracted map (${\rm dT}_{\rm b}$) and brightness temperature contrast ($\delta_{{\rm T}_{\rm b}}$). The middle panels show marks \mref{m0} with $p=2$ and $p=-2$, which select distinct environments and clearly distinguish ionized and neutral regions. The bottom panels show the mark \mref{m1}, which picks out contributions from both the environment in the map, and its corresponding weighted Marked field $\mathbf{M_1}$.}
    \label{fig:combined_brightness_temperature_marks}
\end{figure}

In Figure \ref{fig:combined_brightness_temperature_marks}, we show the simulated brightness-temperature maps together with representative examples of the corresponding marks. For the map shown here, the mass-averaged neutral fraction is $\sim 0.5$. At this stage of reionization, one expects the fluctuation amplitude to be close to its maximum for a Gaussian field, while the actual 21-cm signal remains strongly shaped by the non-Gaussian morphology of ionized and neutral regions. The inhomogeneous distribution of neutral hydrogen therefore gives rise to the coexistence of distinct environments in the IGM, including extended ionized regions and dense neutral structures. We find that the marked maps broadly preserve the morphology of the underlying signal while selectively enhancing different environments. The top row shows different versions of the brightness-temperature field. The second row illustrates how the mark \mref{m0} can be used to enhance ionized and neutral regions separately within the same underlying map. In the middle-left panel, ionized regions are up-weighted and therefore stand out more clearly, whereas in the middle-right panel neutral regions are preferentially enhanced. The bottom-left panel shows the mark \mref{m1}, which enhances the contribution from the local environment of a point irrespective of whether that environment is ionized or neutral. As a result, both types of regions are up-weighted in the transformed map.

The mark can also be used as a weight assigned multiplicatively to the original unmarked field, which in the present case is the mean-subtracted brightness-temperature map, to construct the \textit{marked field}. In this work, we denote the marked field corresponding to the $i^{\rm th}$ mark $m_i$ by $\mathbf{M}_i$. The bottom-right panel of Figure \ref{fig:combined_brightness_temperature_marks} shows a two-dimensional slice of the marked field $\mathbf{M}_1$. The enhanced contrast of the fluctuations is visually apparent, reflecting the additional weighting of each point by its local environment.

\subsection{Marked summary statistics}

To characterize the scale-dependent statistical properties of the EoR 21-cm signal, we use the power spectrum. For a statistically homogeneous field, the power spectrum is defined through the Fourier-space two-point function. For the \HI\ 21-cm brightness-temperature field, it is given by
\begin{equation} \label{power_spectrum}
    \langle \tilde{\delta {\rm T}}_{\rm b}(\mathbf{k})\tilde{\delta {\rm T}}_{\rm b} ^* (\mathbf{k})\rangle = {\rm V} \, {\rm P}({k}),
\end{equation}
where $\tilde{\delta{\rm T}}_{\rm b}(\mathbf{k})$ denotes the Fourier transform of the brightness-temperature field and ${\rm V}$ is the simulation volume. From the simulated maps, we estimate the spherically averaged binned power spectrum and express it in the dimensionless form
\begin{equation}\label{normalized_power_spectrum}
    \texttt{PS}({k}) = \frac{{k}^3}{2\pi^2}\, {\rm P}({k}).
\end{equation}

We use marked power spectra to probe part of the non-Gaussian information contained in the EoR \HI\ 21-cm signal and compare their performance with that of the standard power spectrum. The marked power spectrum is the Fourier-space counterpart of the marked correlation function and may be constructed either from the mark itself or from the marked field. In this work, we consider both the power spectrum of the mark (\textit{Mark Power Spectrum}) and that of the marked field (\textit{Marked Field Power Spectrum}), defined respectively as
\begin{equation} \label{def_MPS}
    \langle \tilde{m}(\mathbf{k}) \, {\tilde{m} }^* (\mathbf{k})\rangle = {\rm V} \, {\text{MPS}}({k}),
\end{equation}
and
\begin{equation} \label{def_MFPS}
    \langle \mathbf{\tilde{M}}(\mathbf{k}) \, {\mathbf{\tilde{M}}} ^* (\mathbf{k})\rangle = {\rm V} \, \text{MFPS}({k}),
\end{equation}
where $\tilde{m}(\mathbf{k})$ is the Fourier transform of the mark $m(\mathbf{x})$ constructed from the \HI\ 21-cm brightness-temperature field, and $\mathbf{\tilde{M}}(\mathbf{k})$ is the Fourier transform of the marked field, defined as $\mathbf{M}(\mathbf{x}) = m(\mathbf{x}) \times {\rm dT}_{\rm b}(\mathbf{x})$. An important practical point is that the same pipeline used to estimate the standard power spectrum can be extended to these marked power spectra with only minimal modification. We further define the dimensionless forms
\begin{equation}\label{normalized_mark_power_spectrum}
    \texttt{MPS}({k}) = \frac{{k}^3}{2\pi^2} \, {\rm MPS}({k}),
\end{equation}
and
\begin{equation}\label{normalized_marked_field_power_spectrum}
    \texttt{MFPS}({k}) = \frac{{k}^3}{2\pi^2}  \, {\rm MFPS}({k}),
\end{equation}
in direct analogy with \texttt{PS}.

It is useful to emphasize that the \textit{Mark Power Spectrum} and the \textit{Marked Field Power Spectrum} are both specific realizations of the broader \textit{Marked Power Spectrum} framework. Throughout this paper, we use both statistics, and denote their dimensionless forms generically by $\Delta^2(k)$ whenever the discussion applies equally to \texttt{PS}, \texttt{MPS}, and \texttt{MFPS}.

\section{Results}

\subsection{The 21-cm Marked Power Spectrum}

\begin{figure}[htbp] 
    \centering
    \includegraphics[width=1\linewidth]{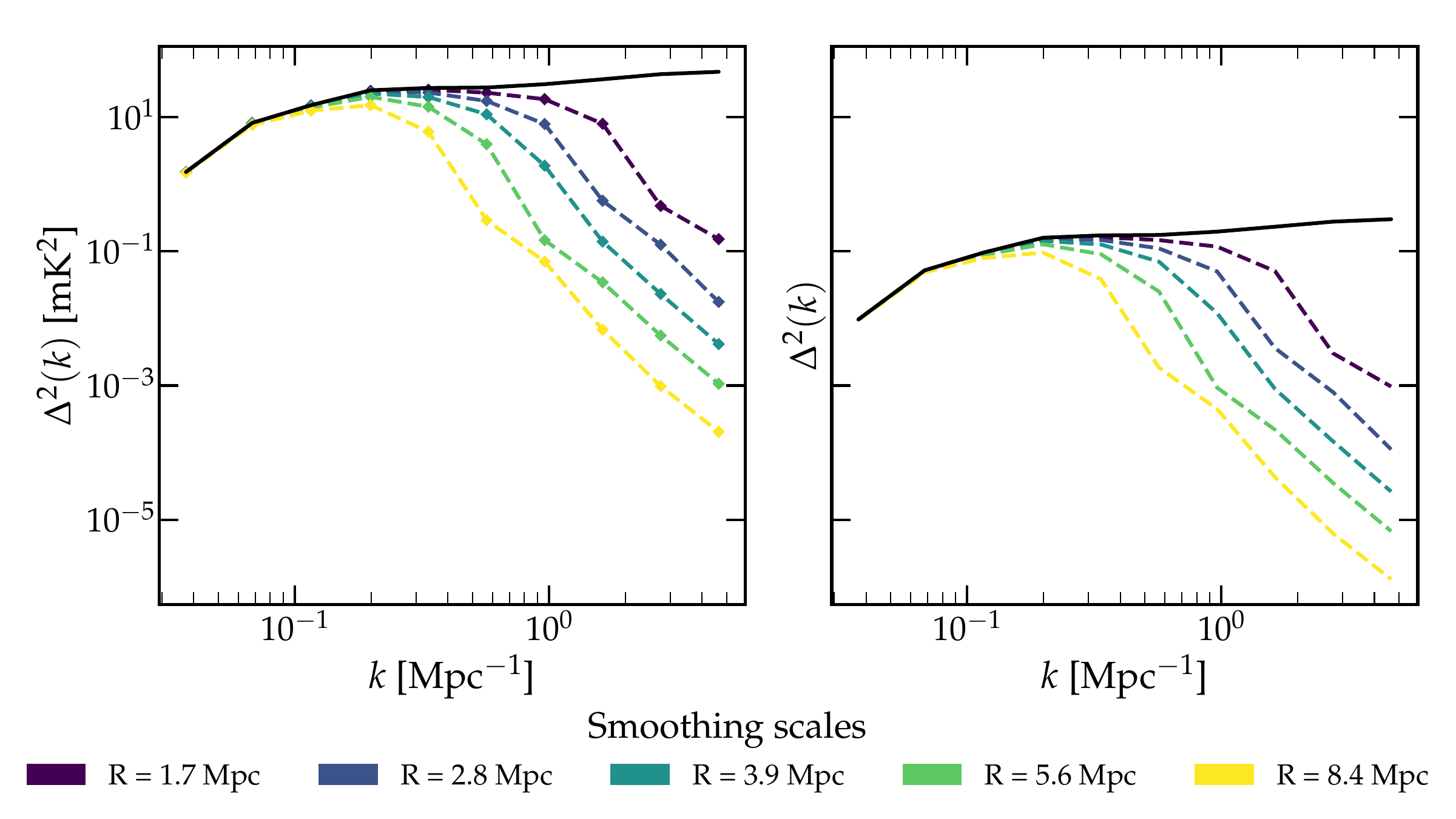}
    \caption{ The power spectrum $\texttt{PS}$ of the versions brightness temperature fields and their smoothed fields. The left panel shows $\texttt{PS}$ of brightness temperature map, solid black line, and smoothed maps which are smoothed using a spherical filter of radius ${\rm R}$, dashed colored lines. The right plot shows $\texttt{PS}$ of brightness temperature contrast, solid colored lines and the corresponding smoothed maps, using the same spherical filter.}   
    \label{fig:smoothed_PS.pdf}
\end{figure}

In this section, we demonstrate how the 21-cm marked power spectrum responds to contributions from different environments in the IGM. Before discussing this in detail, we first compare the power spectra of the differential brightness-temperature maps with those of their smoothed counterparts. Figure \ref{fig:smoothed_PS.pdf} shows \texttt{PS} for the differential brightness-temperature field after smoothing with a spherical filter of radius ${\rm R}$. As expected, smoothing suppresses fluctuations below the corresponding length-scale, which leads to a damping of power at sufficiently high $k$-modes. This suppression becomes progressively stronger at lower $k$ as the smoothing radius increases.

\begin{figure}[htbp] 
    \centering
    \includegraphics[width=1\linewidth]{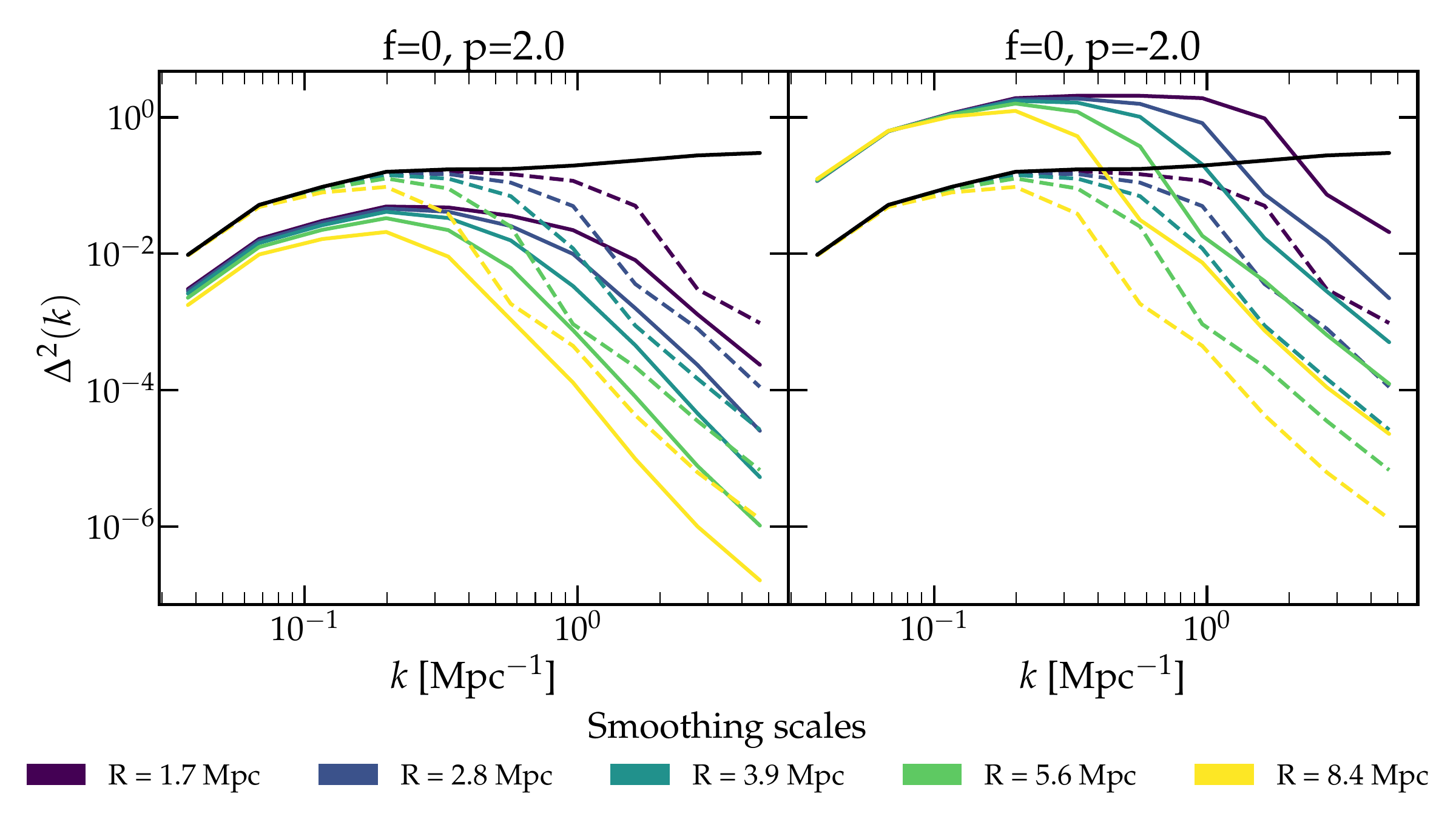}
    \caption{The mark power spectrum (\texttt{MPS}) of the 21-cm differential brightness-temperature field for the mark \mref{m0}, shown as solid coloured lines, for a fixed set of mark parameters at $z=7.52$. The solid black line shows the standard power spectrum (\texttt{PS}) of the brightness-temperature contrast $\delta {\rm T}_{\rm b}$, while the coloured dashed lines show the power spectra of the corresponding smoothed maps for different smoothing radii. The left panel corresponds to \mref{m0}$(f=0,p=2)$, which emphasizes ionized regions, and the right panel corresponds to \mref{m0}$(f=0,p=-2)$, which emphasizes neutral regions.}
    \label{Mark_PS_x_0.5.pdf}
\end{figure}

We next compare \texttt{PS} and \texttt{MPS} for the mark \mref{m0} at different smoothing scales, as shown in Figure \ref{Mark_PS_x_0.5.pdf}. The figure shows that \texttt{MPS} is sensitive to the contribution from different environments. The left panel compares \texttt{PS} (solid black line) and \texttt{MPS} (solid coloured lines) for a mark that emphasizes ionized regions [\mref{m0}$(f=0, p=2)$]. Since ionized regions are less strongly clustered, they contribute relatively less power, leading to a suppression of \texttt{MPS} relative to the standard power spectrum across the range of $k$ shown. The right panel, by contrast, compares \texttt{PS} and \texttt{MPS} for a mark that emphasizes neutral regions [\mref{m0}$(f=0, p=-2)$]. Neutral regions are more strongly clustered and therefore contribute more power at all $k$-modes, resulting in an enhancement of \texttt{MPS} relative to \texttt{PS}. In both cases, the shape of \texttt{MPS} broadly follows that of the power spectrum of the corresponding smoothed map (coloured dashed lines), which is expected because the mark is itself constructed from the smoothed field.

\begin{figure}[htbp]
    \centering
    \includegraphics[width=1\linewidth]{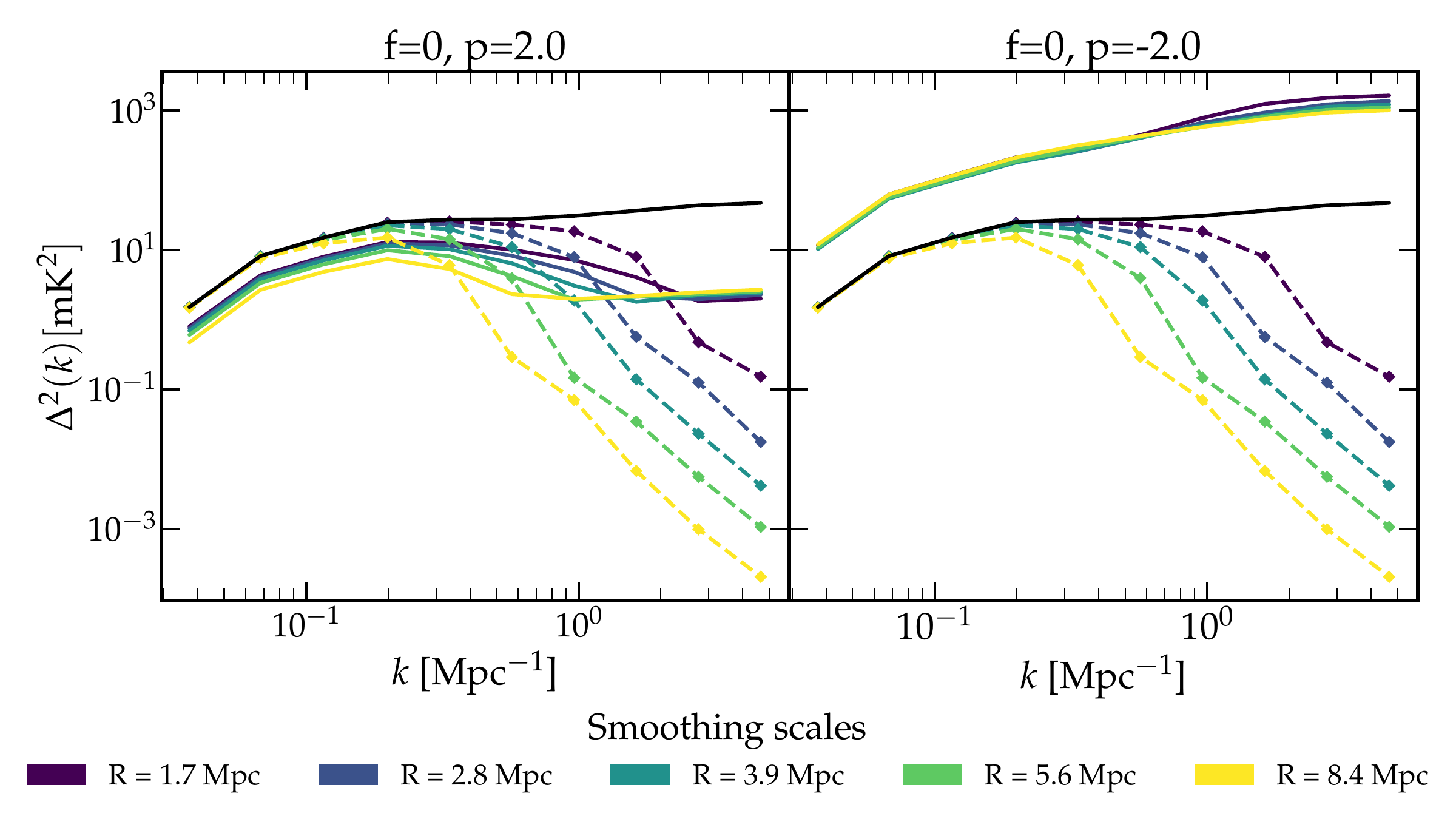}
    \caption{The marked field power spectrum (\texttt{MFPS}) of the 21-cm differential brightness temperature field for the mark \mref{m0}, shown as solid colored lines, for a fixed set of parameters at $z = 7.52$. The solid black line shows the power spectrum (\texttt{PS}) of the mean-subtracted 21-cm differential brightness temperature map ${\rm dT}_{\rm b}$, and the colored dashed lines show the power spectra of the smoothed mean-subtracted map, for different smoothing radii.}
    \label{Marked_PS_x_0.5_m0XdT.pdf}
\end{figure}

Figure \ref{Marked_PS_x_0.5_m0XdT.pdf} shows \texttt{MFPS} (solid coloured lines), together with the standard power spectrum \texttt{PS} of the brightness-temperature field $\delta {\rm T}_{\rm b}(\mathbf{x},z)$ (solid black line) and the power spectra of its smoothed versions (coloured dashed lines). Here, the mark \mref{m0} is applied multiplicatively to the mean-subtracted field to construct the marked field. Similar to \texttt{MPS}, \texttt{MFPS} is sensitive to contributions from different environments in the map. This is expected, since the marked field inherits the environmental weighting encoded in the mark itself. At the same time, because the marked field is obtained by weighting the brightness-temperature field, the resulting \texttt{MFPS} retains a broad similarity to the shape of the underlying power spectrum, while also reflecting the environmental modulation introduced by the mark.

To quantify this behaviour more clearly, we define a \textit{Normalized Marked Field Power Spectrum} (\texttt{NMFPS}), motivated by the normalization of the marked correlation function by the standard two-point correlation function \cite{sheth2005markedcorrelationsgalaxyformation}. We define

\begin{equation}
\label{NMFPS}
    \texttt{NMFPS}(k) = \frac{\texttt{MFPS}(k)}{\texttt{PS}(k)} .
\end{equation}

This estimator quantifies the clustering in the marked field in units of the standard power spectrum. In other words, it measures the \emph{relative environmental clustering} induced by the mark.


\begin{figure}[htbp]
    \centering
    \includegraphics[width=0.7\linewidth]{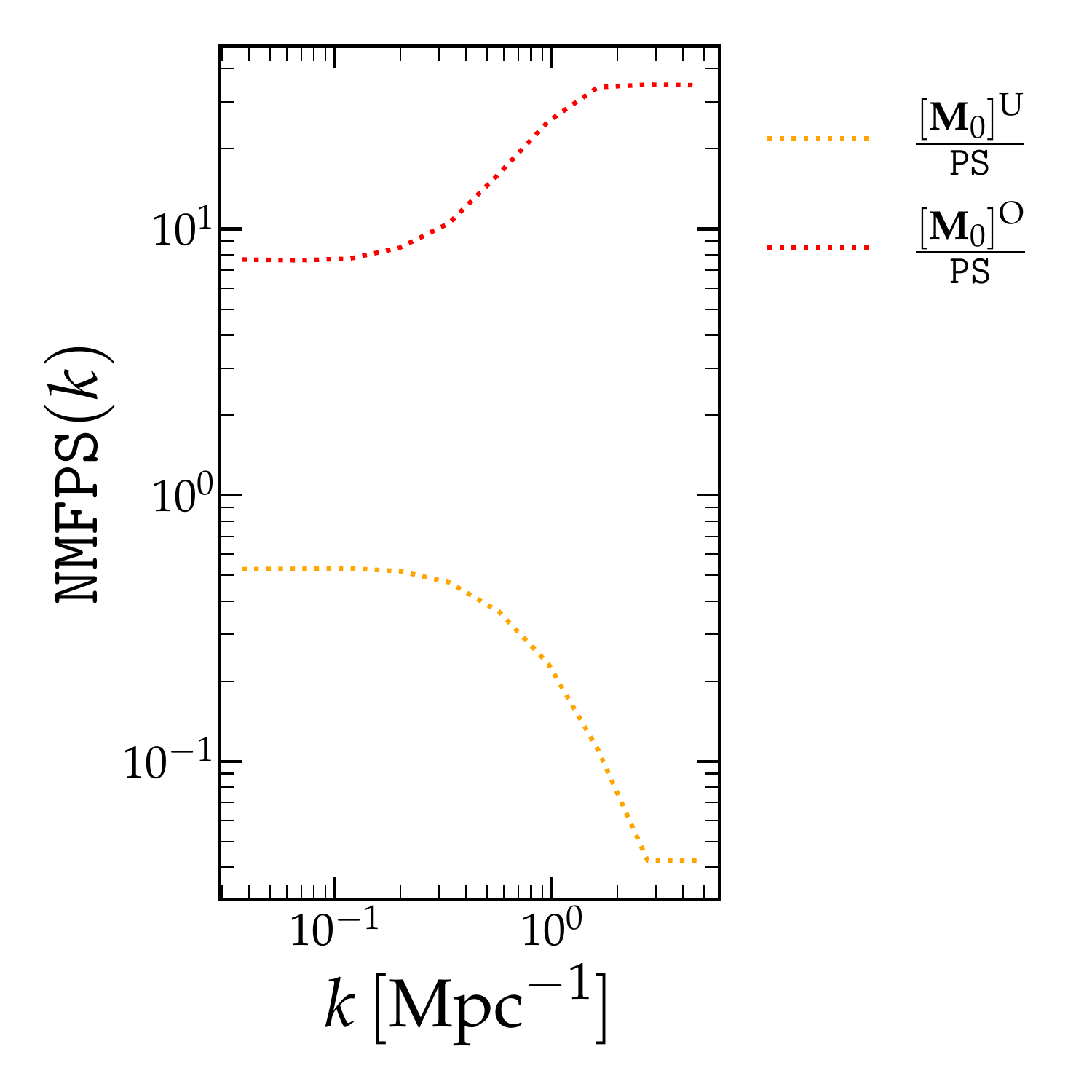}
    \caption{Normalized Marked Field Power Spectrum, $\texttt{NMFPS}(k)$, shown as 
    the ratio $\texttt{MFPS}/\texttt{PS}$. The ratio remains approximately constant 
    over most $k$-modes and deviates at higher $k$, indicating scale-dependent 
    environmental modulation introduced by the mark. The notation $[\mathbf{M}_0]^{\rm U}$ and $[\mathbf{M}_0]^{\rm O}$ represents \texttt{MFPS} for mark \mref{m0}$({\rm R} = 1.68 \, {\rm Mpc}, f=0, p=2)$ and \mref{m0}$({\rm R} = 1.68 \, {\rm Mpc}, f=0, p=-2)$ respectively-emphasizing on the complementary environments in the IGM.}
    \label{fig:NMFPS}
\end{figure}

Figure \ref{fig:NMFPS} shows the behavior of $\texttt{NMFPS}$ as a function of $k$. For most $k$-modes, the ratio remains approximately constant, indicating that the mark primarily rescales clustering over these scales. However, at higher $k$, we observe clear deviations from constant. This change at large $k$-modes (small length scales) again shows that ionized and neutral regions contribute differently depending on their clustering. The deviation of $\texttt{NMFPS}$ from a constant value therefore demonstrates that the marked statistic encodes \emph{scale-dependent environmental bias}. More importantly, since $\texttt{MFPS} = \texttt{NMFPS} \times \texttt{PS}$, it contains both the standard clustering information and the scale-dependent environmental modulation introduced by the mark. The marked power spectrum, therefore, captures both baseline clustering and environment-sensitive structure by isolating the relative environmental contribution through the marking process.

\begin{figure}[htbp]
    \centering
    \includegraphics[width=1\linewidth]{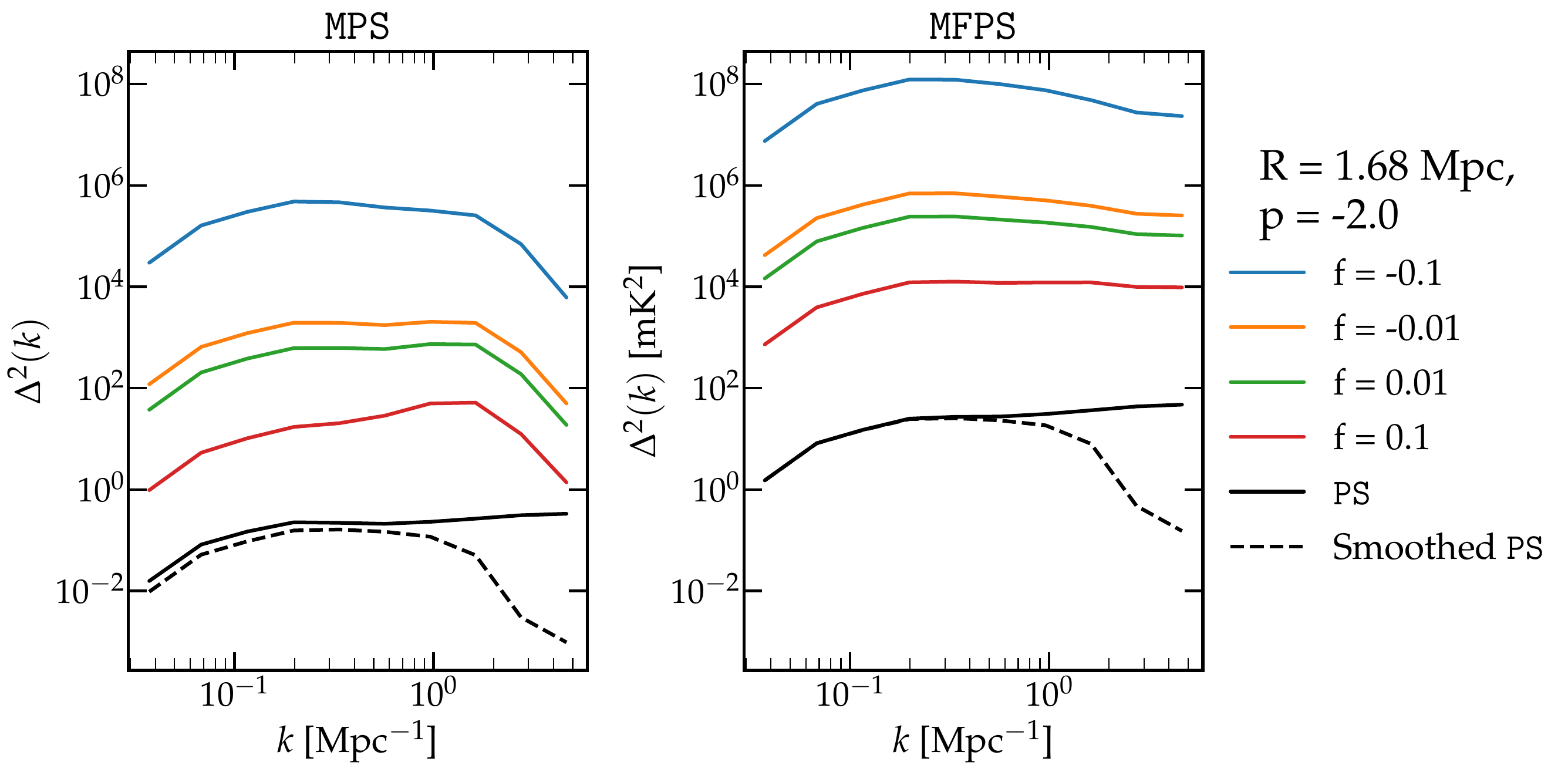}
    \caption{Trend in \texttt{MPS} and \texttt{MFPS} of the HI 21-cm map using mark \mref{m1} for various $f$ values and comparison with \texttt{PS} and power spectrum of the smoothed map-smoothed \texttt{PS} at radius ${\rm R} = 1.68 \, {\rm Mpc}$. In the left plot, the black solid and dashed lines show the power spectra of brightness-temperature contrast $\delta_{{\rm T}_{\rm b}}$. The right plot shows the power spectrum of the mean-subtracted brightness-temperature map ${\rm dT}_{\rm b}$, enabling a concrete comparison of different statistics on the same units.}
    \label{Combined_Power_spectrum_R_03_M3_compare_-2.0.pdf}
\end{figure}
\begin{figure}[htbp]
    \centering
    \includegraphics[width=1\linewidth]{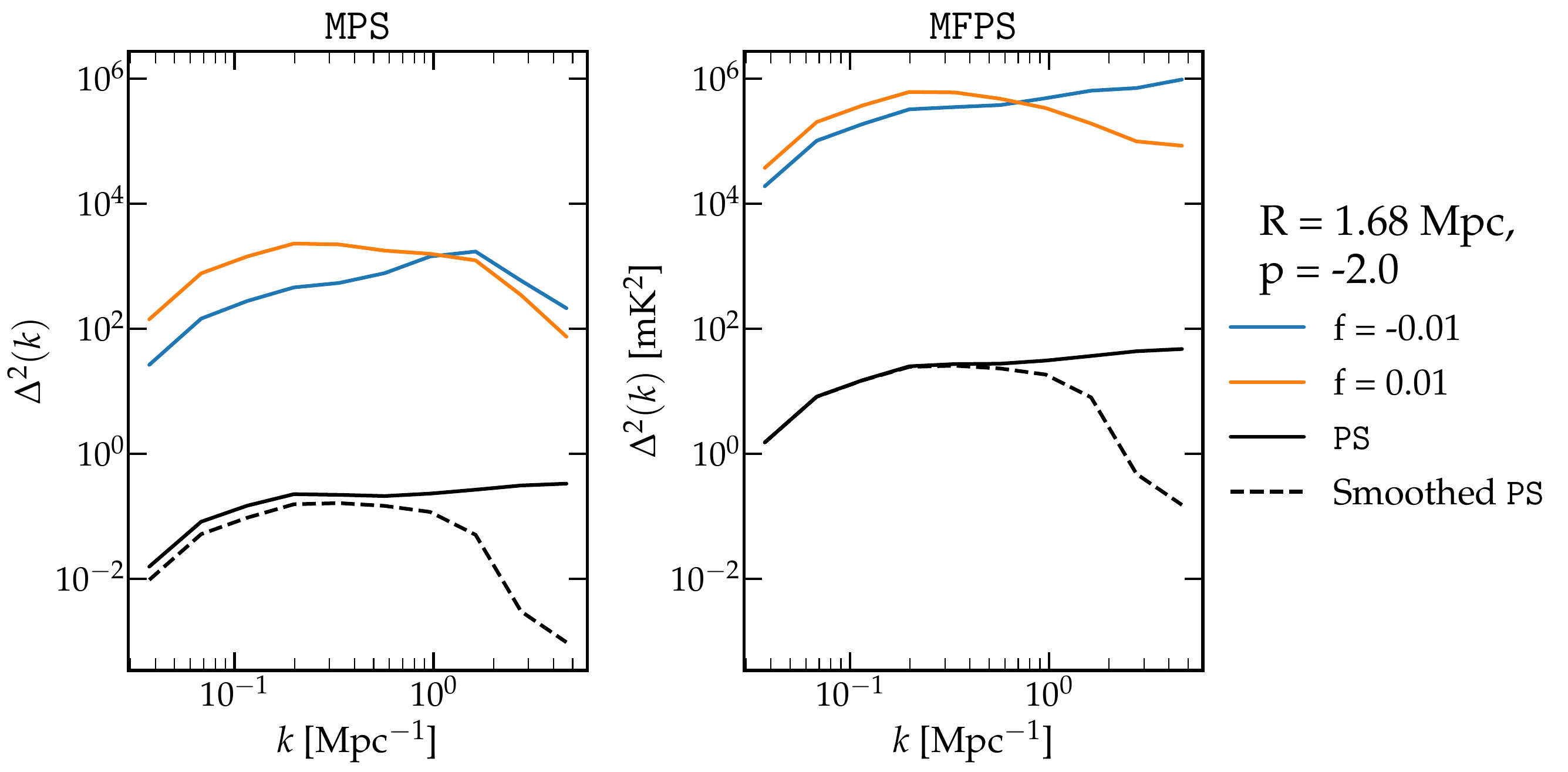}
    \caption{Same as Figure \ref{Combined_Power_spectrum_R_03_M3_compare_-2.0.pdf} using mark \mref{m2}. }
    \label{Combined_Power_spectrum_R_03_M5_compare_-2.0.pdf}
\end{figure}

Next, we show \texttt{MPS} and \texttt{MFPS} for marks \mref{m1} and \mref{m2}. Unlike \mref{m0}, for \mref{m1} and \mref{m2} the thresholding parameter is weighted by some form of brightness temperature field ${\rm dT}_{\rm b}(\mathbf{x},z)$. We restrict our analysis to negative values of exponent in marks as they yield `unstable' power for positive exponents, as discussed in the next sections, shown in Figure \ref{fig:Combined_Power_spectrum_R_03_M3_compare_2.0.pdf}. Figures \ref{Combined_Power_spectrum_R_03_M3_compare_-2.0.pdf} and \ref{Combined_Power_spectrum_R_03_M5_compare_-2.0.pdf} shows \texttt{MPS} and \texttt{MFPS} for marks \mref{m1}$(f,p=-2)$ and \mref{m2}$(f,p=-2)$ for various thresholding parameter $f$. From Figure \ref{fig:combined_brightness_temperature_marks}, we observe a clear systematic trend in the amplitude of the Marked Power Spectrum across all $k$-modes as the mark parameter $f$ is varied. For positive values of $f$, the overall amplitude of the Marked Power Spectrum decreases, whereas for negative values it increases.



Analogous to \mref{m0}, the shapes of the \texttt{MPS} and \texttt{MFPS} for mark \mref{m1} closely follow those of the smoothed fields and brightness temperature fields, respectively. We also observe that, for certain choices of $f$, different marked power spectra exhibit very similar amplitudes and shapes across all $k$-modes. This points to a clear degeneracy in the mark parameters and the marked power spectra. Such degeneracy arises because similar transformations of the brightness temperature maps yield nearly equivalent reweightings, leading to closely overlapping power spectra. While this degeneracy may be scale-dependent, its dominant effect here appears as overall amplitude similarity rather than strong shape variation. Importantly, the Marked Power Spectrum is sensitive to both: to the environment of emphasis and its degree. Marks such as \mref{m1} and \mref{m2} are defined as general functional forms of the brightness temperature rather than tied to a specific physical prescription, considering negative $f$ values allows us to probe a broader class of transformations. This way. we treat marking as a general mathematical framework for reweighting the 21-cm signal, effectively capturing its statistical properties.

So far, we have shown that the 21-cm marked power spectrum is sensitive to environments selected by the mark parameters for the different functional forms considered here. At the same time, the choice of mark parameters is not unique, since the choice of the mark itself is not unique. As discussed in \cite{Cowell_2024}, marks with underlying symmetries can exhibit the property that apparently different parameter choices, or even different functional forms, produce distinct transformations while retaining similar information content. More importantly, the application of a mark does not in itself guarantee that the resulting marked power spectrum will contain additional useful information beyond the standard power spectrum, as will be shown in the following sections. These considerations motivate us to quantify the information content of the 21-cm marked power spectrum explicitly.

\subsection{Information content in the 21-cm Marked Power Spectrum}


To quantify the information content of the 21-cm marked power spectrum, we perform a Fisher-matrix analysis and forecast the EoR model parameters using \texttt{MPS} and \texttt{MFPS} for the various marks introduced in the previous sections, and compare the results with those obtained from the signal power spectrum \texttt{PS}. To compute the derivatives of the estimators entering the Fisher analysis, we generate \HI\ 21-cm maps by varying the EoR parameters around their fiducial values as follows: ${\rm N}_{\rm ion} = 24.71$ ($23.21$--$27.21$, step size $0.5$), ${\rm M}_{\rm h,\,min} = 14 \times 10^8{\rm M}_\odot$ ($10 \times 10^8 \, {\rm M}_{\odot}$--$18 \times 10^8 \, {\rm M}_{\odot}$, step size $10^8 \, {\rm M}_\odot$), and ${\rm R}_{\rm mfp} = 15.00~{\rm Mpc}$ ($8.28$--$21.72$, step size $1.68~{\rm Mpc}$). The marginalized error squared associated with the EoR parameter $\theta_\alpha$ satisfies $\sigma^2(\theta_\alpha) \ge (\mathcal{F}_{\alpha\alpha})^{-1}$, where the Fisher matrix is given by


\begin{equation}
\mathcal{F}_{\alpha \beta}(z) = \sum_{i, j} 
\left[
\frac{\partial \Delta^2_{21}(k_i\, ; \, z)}{\partial q_\alpha} \,
\left[ \mathbf{C}(k_i, k_j\, ; \,z) \right]^{-1} \,
\frac{\partial \Delta^2_{21}(k_j\, ; \, z)}{\partial q_\beta}
\right],
\end{equation}


where $\Delta^2_{21}(k\, ; \, z)$ denotes the dimensionless power spectrum, i.e. \texttt{PS}, \texttt{MPS}, or \texttt{MFPS}, and $\mathbf{C}(k_i,k_j \, ; \, z)$ is the covariance matrix between the $k$-modes at redshift $z$ for the corresponding estimator. For the EoR 21-cm signal, the errors are in general expected to be correlated across $k$-modes, particularly on small scales, and to depend on the ionization state of the map \cite{Mondal_2015, Mondal_2015_2, Mondal_2016}. In the present work, however, we adopt a simplified treatment and assume Gaussian errors that are uncorrelated between different $k$-bins. Within this approximation, a smaller marginalized error corresponds to a stronger constraining power of the estimator. To evaluate the Fisher matrix $\mathcal{F}_{\alpha\beta}$, we use the estimator covariance $\mathbf{C}_{ij}$ together with the derivatives $\partial \Delta^2(k)/\partial q_\alpha$ of the estimators (\texttt{PS}, \texttt{MPS}, and \texttt{MFPS}) with respect to the EoR parameters. These derivatives are computed numerically using a six-point stencil method, with parameter increments set by the grid of values around the fiducial model listed above. For the present analysis, we restrict the Fisher summation to seven logarithmically spaced $k$-bins in the range $0.12~{\rm Mpc}^{-1}$ to $2.75~{\rm Mpc}^{-1}$. Moreover, for all marks we fix the smoothing scale to ${\rm R}=1.68~{\rm Mpc}$ while varying the other mark parameters.

\begin{table}[htbp!] 
\centering
\begin{tabular}{| c | c | c | c |}
\hline
\textbf{Mark} & \textbf{Mark parameters} & \texttt{MPS} & \texttt{MFPS} \\
\hline
\multirow{2}{*}{\mref{m0}} 
 & f=0.0, p=2  & $[{m_0}]^{\rm U}$ & $[{\mathbf{M}_0}]^{\rm U}$ \\
 & f=0.0, p=-2  & $[{m_0}]^{\rm O}$ & $[{\mathbf{M}_0}]^{\rm O}$ \\
\hline
\multirow{3}{*}{\mref{m1}} 
 & f=0.1, p=-2   & $[{m_1}]^{\rm A}$ & $[{\mathbf{M}_1}]^{\rm A}$ \\
 & f=0.01, p=-2  & $[{m_1}]^{\rm B}$ & $[{\mathbf{M}_1}]^{\rm B}$ \\
 & f=-0.01, p=-2 & $[{m_1}]^{\rm C}$ & $[{\mathbf{M}_1}]^{\rm C}$ \\
  & f=0.01, p=-1  & $[{m_1}]^{\rm D}$ & -- \\
 
\hline
\multirow{2}{*}{\mref{m2}} 
 & f=0.01, p=-2  & $[{m_2}]^{\rm B}$ & $[{\mathbf{M}_2}]^{\rm B}$ \\
 & f=-0.01, p=-2 & $[{m_2}]^{\rm C}$ & $[{\mathbf{M}_2}]^{\rm C}$ \\
  & f=0.01, p=-1  & $[{m_2}]^{\rm D}$ & -- \\
\hline
\end{tabular}
\caption{Various Marked Power Spectra, at fixed smoothing scale ${\rm R = 1.68 \, Mpc}$, and their notations. The letters A, B, C, D, U, and O in the notation of \texttt{MPS} or \texttt{MFPS} indicate a specific choice of mark parameters for any mark used. Particularly, ``U'' and ``O'' are used to denote marks that are informed of ionized and neutral regions, respectively, which are analogous to ``under'' and ``over'' dense regions in the 21-cm brightness temperature field for mark \mref{m0}.}
\label{Table:Marked Power spectrum and notations}
\end{table}

\begin{figure}[htbp]
    \centering

    \subcaptionbox{Performance of \texttt{MPS} against \texttt{PS} for mark \mref{m0}
    \label{fig_Fisher_mps_m0}}[0.6\linewidth]{%
        \includegraphics[width=\linewidth]{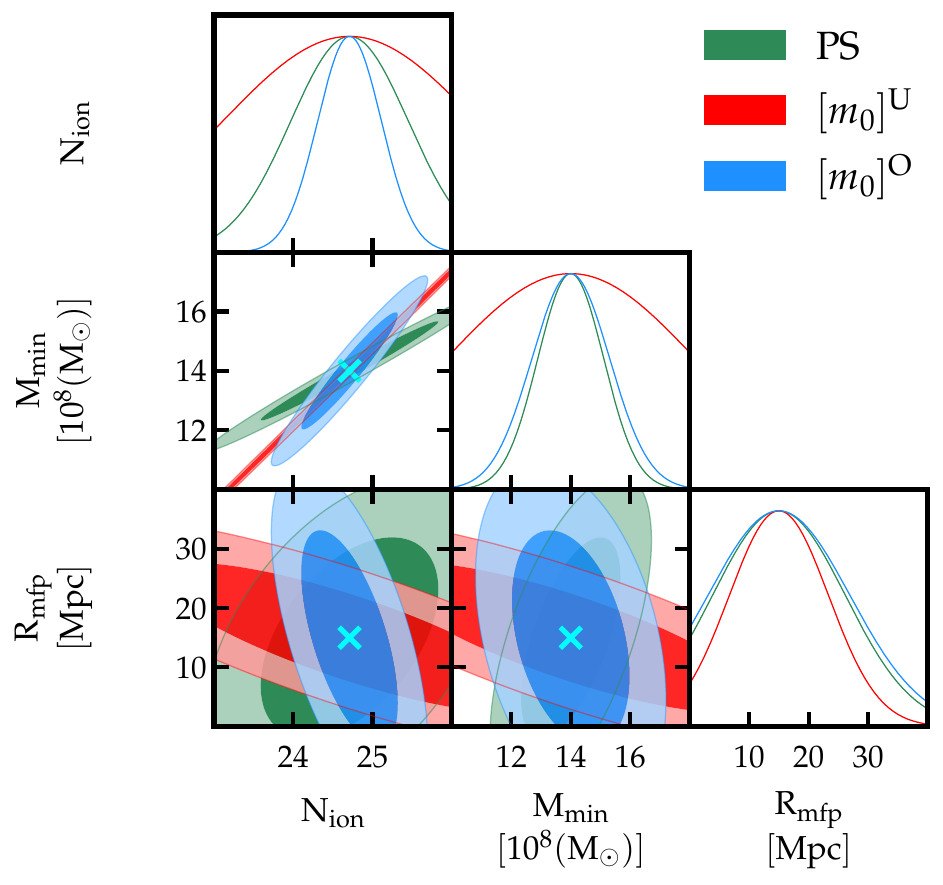}}

    \vspace{0.5cm}

    \subcaptionbox{Performance of \texttt{MPS} against \texttt{PS} for mark \mref{m1}
    \label{fig_Fisher_mps_m3}}[0.45\linewidth]{%
        \includegraphics[width=\linewidth]{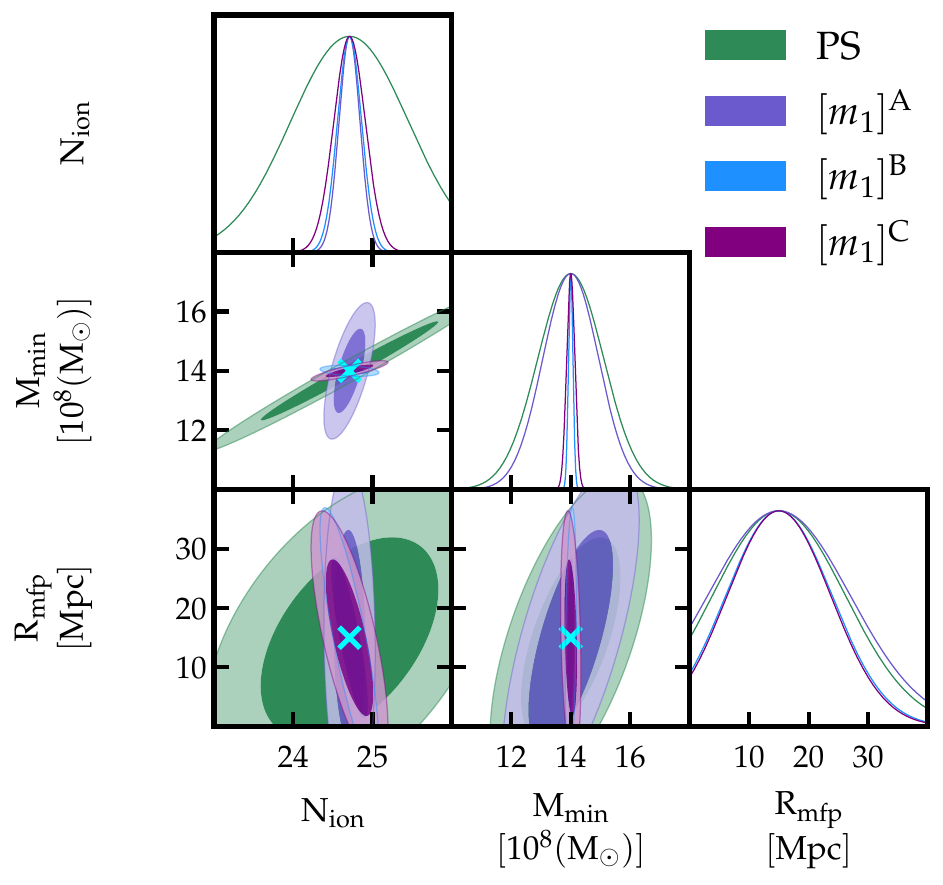}}
    \hfill
    \subcaptionbox{Performance of \texttt{MPS} against \texttt{PS} for mark \mref{m2}
    \label{fig_Fisher_mps_m5}}[0.45\linewidth]{%
        \includegraphics[width=\linewidth]{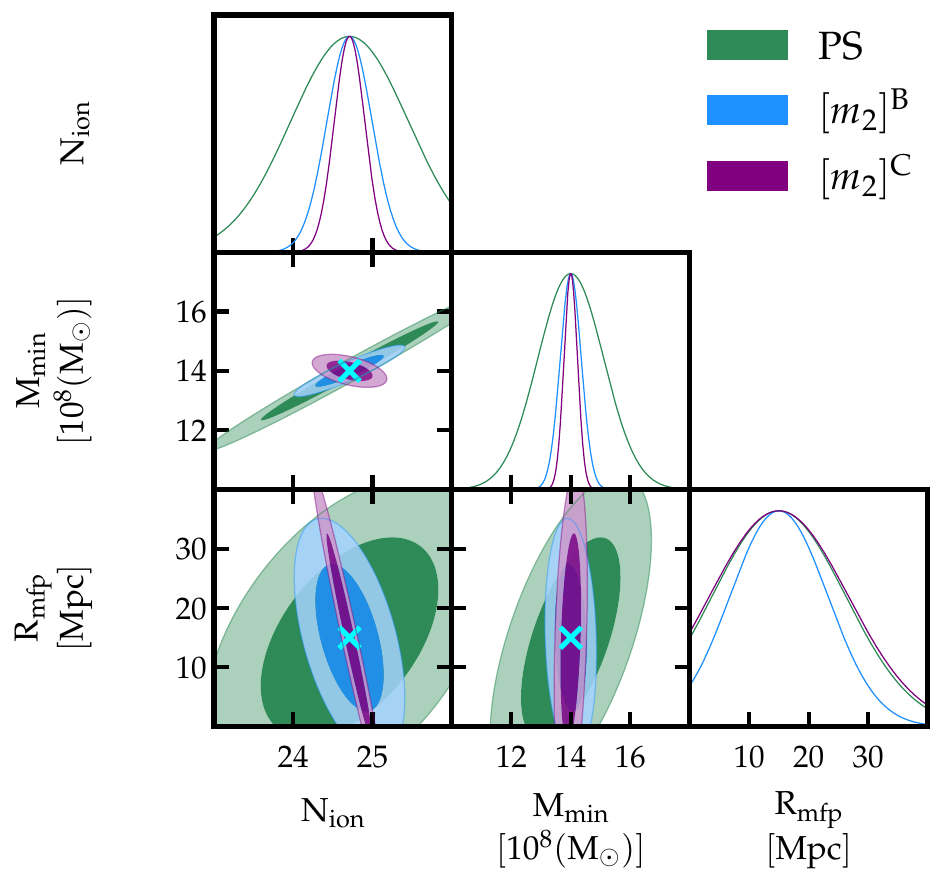}}

    \caption{The 1$\sigma$ - 2$\sigma$ error contours in the Fisher forecast of the EoR model parameters using 21-cm Power Spectrum (\texttt{PS}) and Mark Power Spectrum (\texttt{MPS}) at $z=7.52$ ($\bar{\text{x}}_\text{HI}=0.5$) for marks \mref{m0}, \mref{m1}, and \mref{m2}. The cyan cross in the triangle plot represents the fiducial values of the EoR model parameter.}
    \label{fig_Fisher_MPS_all}
\end{figure}

The comparison of Fisher forecasts of EoR model parameters between \texttt{PS} and \texttt{MPS} is shown in Figure \ref{fig_Fisher_MPS_all}. Figure \ref{fig_Fisher_mps_m0} shows the triangle plot of parameter forecasts using the 21-cm power spectrum (\texttt{PS}) and marked power spectrum (\texttt{MPS}) constructed with \mref{m0}. In contrast, the bottom panels show the corresponding forecasts using \mref{m1} and \mref{m2}. For these marks, \texttt{MPS} systematically improves forecast compared to \texttt{PS}. In particular, the degeneracy between ${\rm N}_{\rm ion}$ and ${\rm R}_{\rm mfp}$ is reduced to a noticeable degree.

The improvement manifests differently depending on the nature of the mark. The simple and physically motivated structure of \mref{m0} allows construction of complementary estimators, $[m_0]^{\rm U}$ and $[m_0]^{\rm O}$, which selectively emphasize ionized and neutral regions, respectively. This selective environmental weighting leads to a visible rotation of the degeneracy direction in the ${\rm N}_{\rm ion}$--${\rm R}_{\rm mfp}$ plane, indicating that complementary information is being accessed. In contrast, the more general functional forms \mref{m1} and \mref{m2}, designed to capture non-Gaussian features from all environments simultaneously, mainly lead to a shrinkage of the error contours, while breaking the EoR model parameter degeneracy to some degree. In a sense, these marks boost contributions from both ionized and neutral regions, resulting in a better forecast of the EoR model parameters, particularly for ${\rm N}_{\rm ion}$ and ${\rm M}_{\rm min}$, while improvement in ${\rm R}_{\rm mfp}$ remains moderate due to its intrinsic degeneracy \cite{Tiwari_2022}. 

\begin{figure}[htbp]
    \centering

    \subcaptionbox{Performance of \texttt{MFPS} against \texttt{PS} for mark \mref{m0}
    \label{fig_Fisher_mfps_m0}}[0.6\linewidth]{%
        \includegraphics[width=\linewidth]{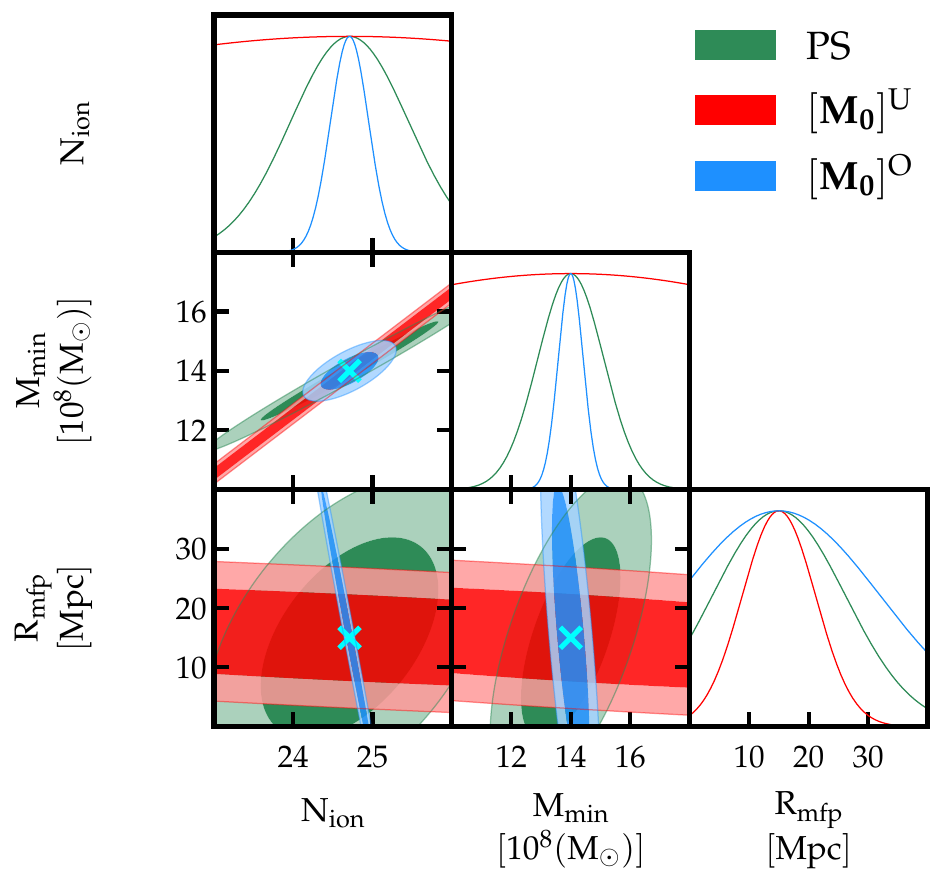}}

    \vspace{0.5cm}

    \subcaptionbox{Performance of \texttt{MFPS} against \texttt{PS} for mark \mref{m1}
    \label{fig_Fisher_mfps_m3}}[0.45\linewidth]{%
        \includegraphics[width=\linewidth]{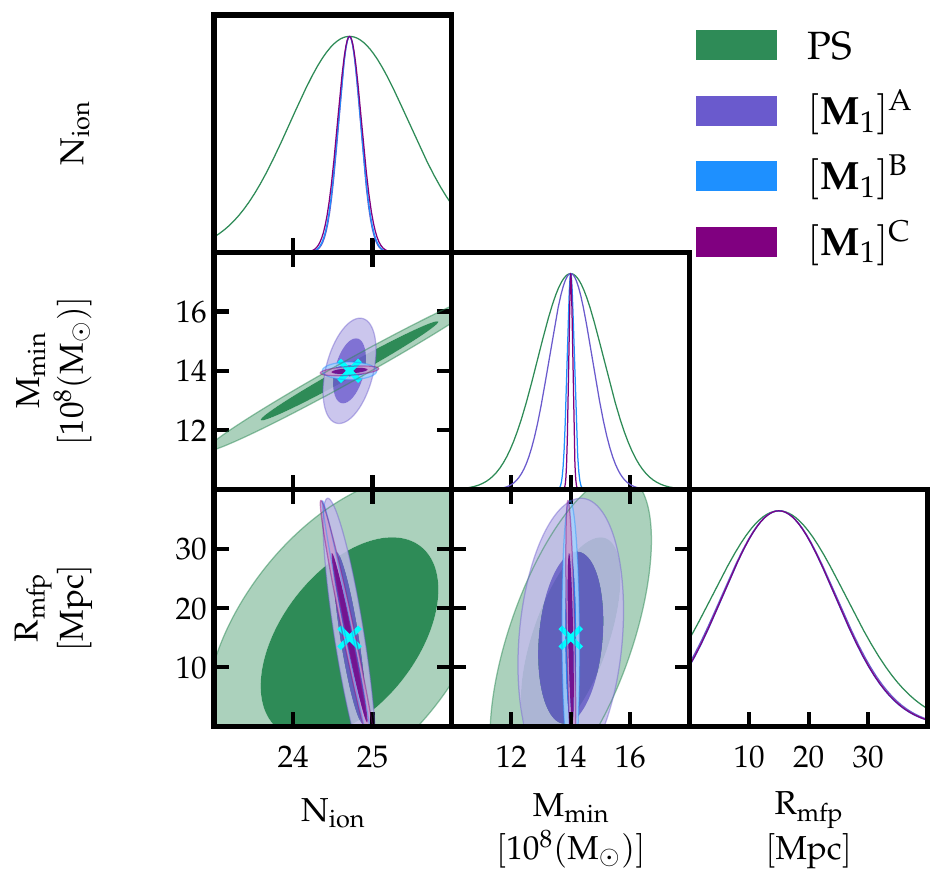}}
    \hfill
    \subcaptionbox{Performance of \texttt{MFPS} against \texttt{PS} for mark \mref{m2}
    \label{fig_Fisher_mfps_m5}}[0.45\linewidth]{%
        \includegraphics[width=\linewidth]{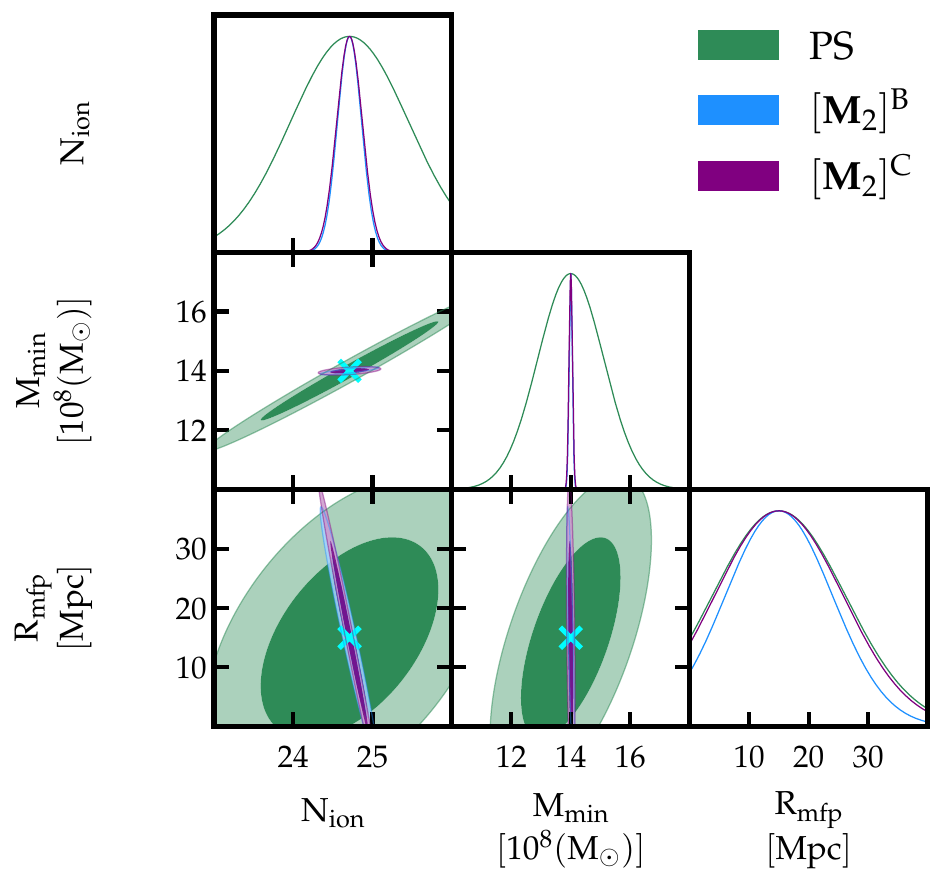}}

    \caption{The 1$\sigma$ - 2$\sigma$ error contours in the Fisher forecast of the EoR model parameters using 21-cm Power Spectrum (\texttt{PS}) and Mark Field Power Spectrum (\texttt{MFPS}) at $z=7.52$ ($\bar{\text{x}}_\text{HI}=0.5$) for marks \mref{m0}, \mref{m1}, and \mref{m2}. The cyan cross in the triangle plot represents the fiducial values of the EoR model parameter.}
    \label{fig_Fisher_MFPS_all}
\end{figure}

\begin{figure}[htbp]
    \centering
    \includegraphics[width=1\linewidth]{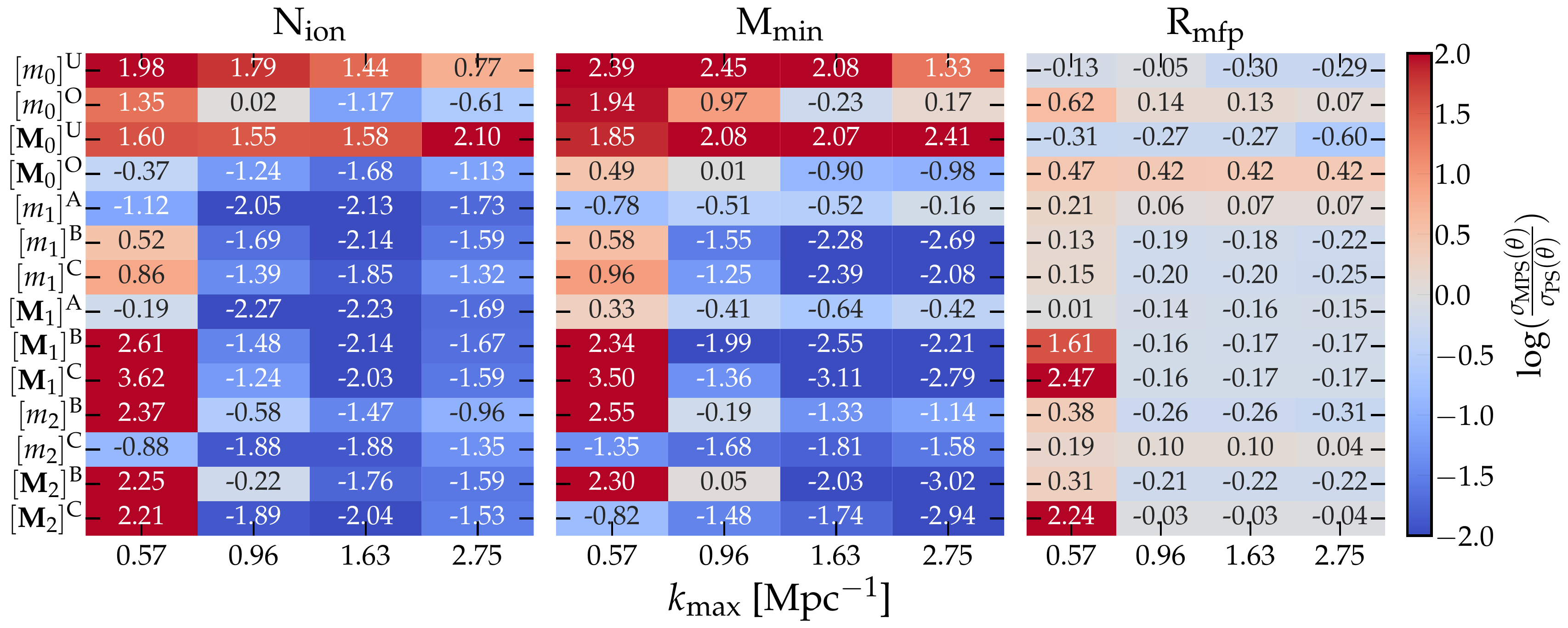}
    \caption{The log of relative marginalized error in the EoR parameters of various Marked power spectrum as a function of maximum $k$-mode ($k_{\rm max} \, [{\rm Mpc}^{-1}])$ included in the fisher metric analysis. The blue boxes shows improvements in the forecast complementary to the red colored boxes. For some marks, this improvement is achieved at the cost of information loss in other degenerate EoR parameter.}
    \label{fig:Marginalized_error2.pdf}
\end{figure}

Figure \ref{fig_Fisher_MFPS_all} shows the analogous comparison to Figure \ref{fig_Fisher_mps_m0}, using \texttt{MFPS}. The correlation of a point and its local environment makes \texttt{MFPS} an effective provider of higher-order contributions, as it measures the power spectrum of a non-linearly transformed field. Compared to \texttt{PS}, \texttt{MFPS} consistently provides better forecast for these stable marks considered here, with improvements primarily arising from a stronger shrinkage of the error contours.
We now also compare the performance of various marks through their relative marginalized errors, shown in Figure \ref{fig:Marginalized_error2.pdf}. The figure presents the relative marginalized errors for each EoR model parameter $\theta$ obtained from $k$-modes with $k < 2.75\,{\rm Mpc}^{-1}$ and their combinations, for different \texttt{MPS} and \texttt{MFPS} estimators. The complementary nature of $[m_0]^{\rm U}$ and $[m_0]^{\rm O}$ is evident: estimators focusing on ionized regions provide improved constraints on ${\rm R}_{\rm mfp}$ at the expense of degradation in ${\rm N}_{\rm ion}$ and ${\rm M}_{\rm min}$, while those focusing on neutral regions improve ${\rm N}_{\rm ion}$ and ${\rm M}_{\rm min}$ with a corresponding mild trade-off in ${\rm R}_{\rm mfp}$. 
Marks such as \mref{m1} and \mref{m2}, constructed to encode information from all environments simultaneously, provide improved forecasts across all EoR parameters. Among the marks considered, \mref{m2} performs slightly better overall than \mref{m1}, reflecting its stronger ability to capture non-Gaussian information. Furthermore, as additional $k$-modes are included in the analysis, the relative marginalized errors decreases as expected. For some marks the improvement continues steadily with increasing $k_{\max}$, while for others it begins to saturate for $k_{\max} \leq 2.75\,{\rm Mpc}^{-1}$. 

Interestingly, we find that negative values of $f$, for marks can outperform their positive counterparts in the Fisher forecasts. Although the overall shape of the Marked Power Spectrum remains similar, negative $f$ values yield systematically larger amplitudes and enhance the relative variation of the spectrum with respect to the EoR model parameters. This increased parameter sensitivity leads to improved forecast, demonstrating that the sign of the environmental response plays a non-trivial role in optimizing the information content of 21-cm marked statistics.


\section{Discussion and Summary}

\subsection{Advantages of the 21-cm Marked Power Spectrum}

In this work, we have explored the Marked Power Spectrum, which provided a conceptually simple yet statistically powerful extension of the standard power spectrum for analyzing the EoR 21-cm signal. Its main strengths lie in its practical simplicity, controlled environmental sensitivity, and ability to extract non-Gaussian information without resorting to the estimation of higher-order statistics.

\begin{itemize}

\item \textbf{Practical implementation and computational efficiency}

The primary advantage of the Marked Power Spectrum is its straightforward implementation. If an inference pipeline already includes a standard power spectrum estimator, one can incorporate the Marked Power Spectrum with a minimal modification. It boils down to defining a suitable mark, transforming or reweighting the map 
and computing the power spectrum of the transformed field. No additional simulation framework, estimator architecture, or statistical machinery beyond two-point analysis is required. Computationally, the cost remains comparable to that of the standard power spectrum, since the only additional operation is a local transformation of the field prior to the Fourier transform. In contrast to explicitly higher-order statistics such as the bispectrum, which are substantially more expensive to compute and model, the Marked Power Spectrum retains the computational simplicity of a two-point estimator while enhancing sensitivity to non-Gaussian structure.

\item \textbf{Controlled environmental sensitivity}

The marked approach introduces explicit free parameters that govern the transformation. These parameters allow systematic control over which 
environments in the map are emphasized or suppressed. For example, suitable functional forms can preferentially weight neutral or ionized regions. This tunability makes the method conceptually transparent. On varying mark parameters, one can visually and statistically alter the relative contribution of different environments and directly quantify how the information content changes. In this sense, the Marked Power Spectrum enables exploration of multiple aspects of the same map without introducing a fundamentally new observable.

\item  \textbf{Forecast improvement and breaking model parameter degeneracy}

For suitably constructed marks, the Marked Power Spectrum outperforms the standard power spectrum in Fisher forecasts of EoR model parameters. The improvement, however, is not universal for arbitrary marks. Identifying effective functional forms for mark is central to the analysis.

Well-designed marks enhance sensitivity to environmental contrasts in the signal, thereby reducing marginalized errors. In some cases, the orientation of the signal model parameter error contours are altered, indicating sensitivity to structurally different morphological features of the EoR. In other cases, the orientation of the error contours remain similar, but constraints shrink due to increased parameter sensitivity. Interestingly, when only a small number of $k$-modes are included, the marked statistic may initially perform worse than the standard power spectrum. However, as progressively smaller scales are incorporated, the marked power spectra's environmental sensitivity leads to a more rapid reduction in parameter uncertainty. Thus, the advantage of marking becomes increasingly pronounced as additional modes are included.

\item  \textbf{Connection to higher-order information}

Although the Marked Power Spectrum is formally a reweighted two-point statistic, the non-linear transformation of the field modifies the distribution of fluctuations. Since the EoR 21-cm signal is intrinsically non-Gaussian, this transformation can redistribute higher-order information into the two-point function of the transformed field, analogous to Gaussianizing the data-set \cite{Neyrinck_2009, Neyrinck_2011}, as mentioned before. 

In this sense, the Marked Power Spectrum acts as a generalized two-point statistic. Rather than directly computing higher-order correlations, the transformation reorganizes non-Gaussian information such that the power spectrum of the transformed field captures aspects of higher-order structure. This explains why the Marked Power Spectrum can improve parameter constraints while maintaining computational efficiency. 

\item  \textbf{Complementarity of multiple marks}

Different marks can emphasize complementary environments within the same map. Since each transformation encodes correlations between a point and its surroundings differently, multiple marked power spectra can, in principle, probe complementary information. However, combining such estimators requires careful treatment of covariance and mode mixing, as information overlap may be substantial. Information gain due to combining several marks are not necessarily additive and must be quantified explicitly.

\end{itemize}

\subsection{Challenges and Limitations}

Despite its advantages, the 21-cm marked power spectrum presents several conceptual and practical challenges.

\begin{figure}[htbp]
    \centering
    \includegraphics[width=1\linewidth]{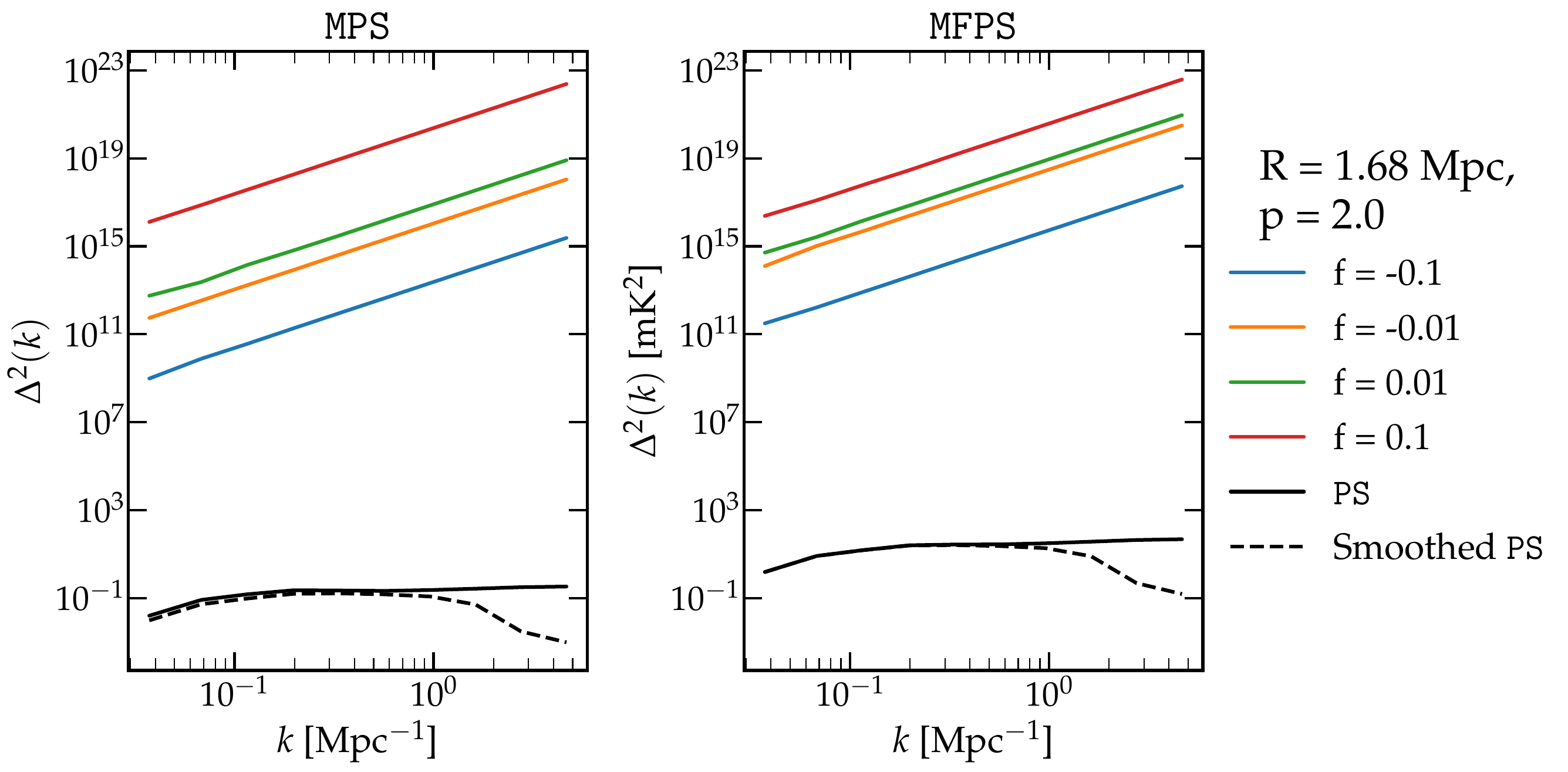}
    \caption{The marked power spectrum for the mark \mref{m1} for different choice of mark parameters which leads unstable rise in the power spectra, leading to loss of valuable information. This demonstrates the importance of choice of mark parameters for the 21-cm marked power spectrum.}
    \label{fig:Combined_Power_spectrum_R_03_M3_compare_2.0.pdf}
\end{figure}

\begin{figure}[htbp]
    \centering
    \subcaptionbox{\label{fig_Fisher_m3_pm12}}[0.45\linewidth]{%
        \includegraphics[width=\linewidth]{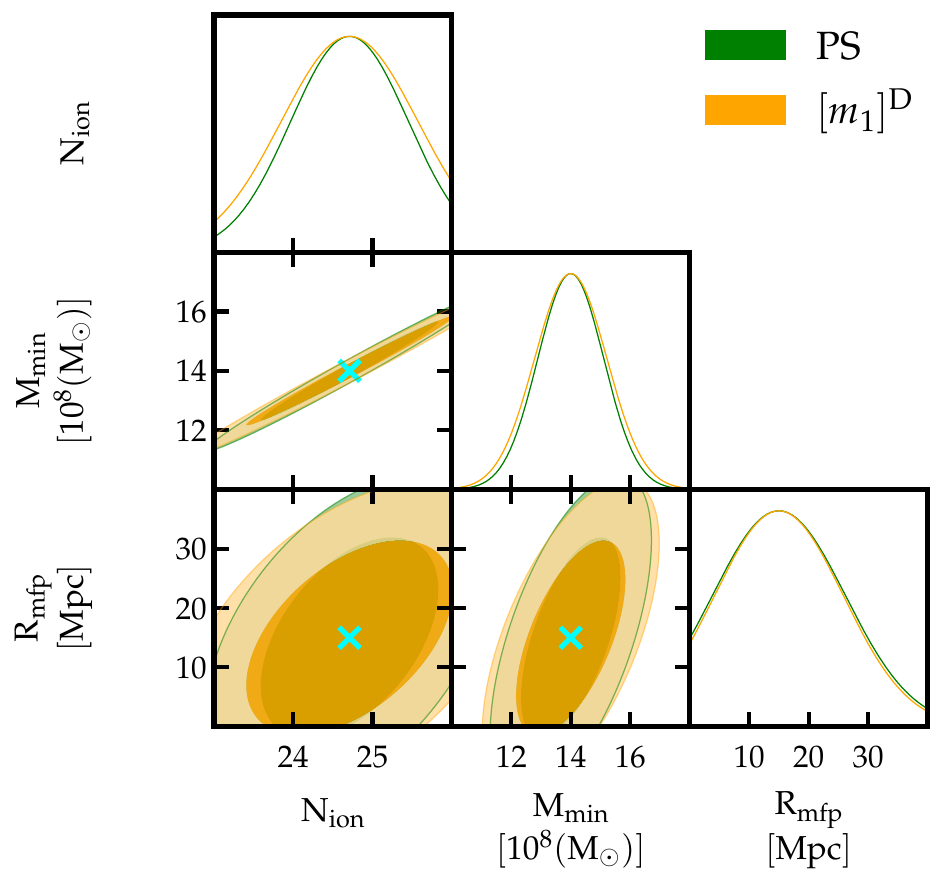}}
    \hfill
    \subcaptionbox{\label{fig_Fisher_m5_pm12}}[0.45\linewidth]{%
        \includegraphics[width=\linewidth]{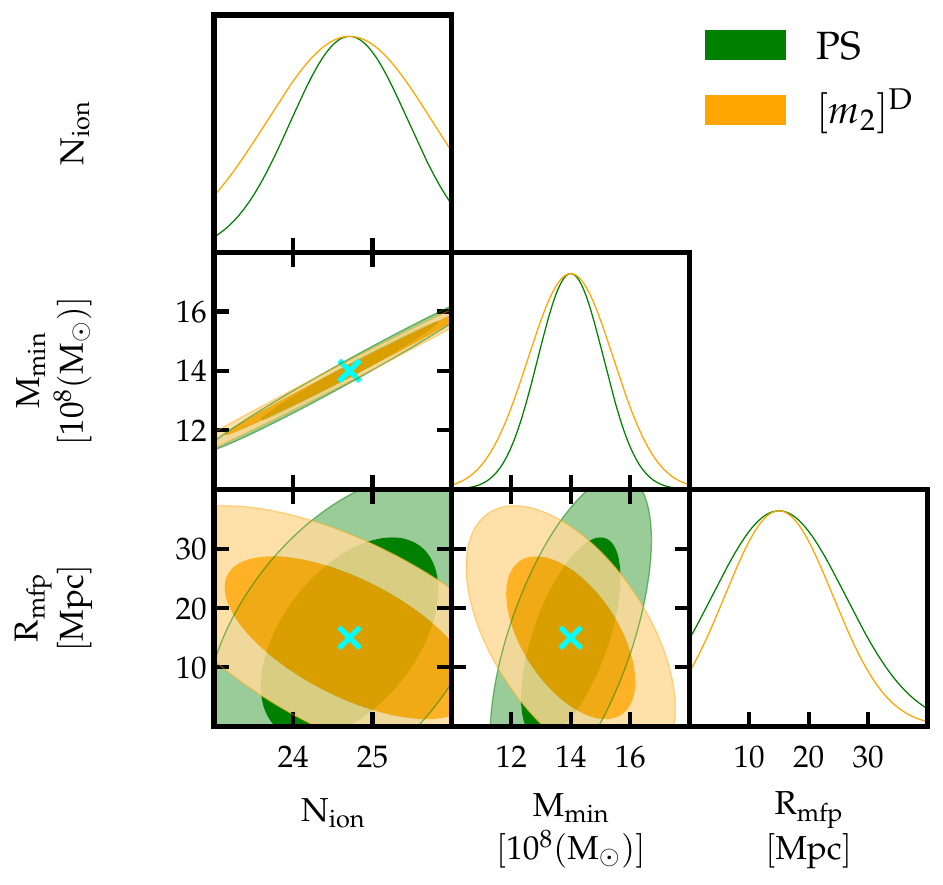}}
    \caption{\textit{Left plot}: The \texttt{MPS}, for mark \mref{m1}, fails to capture any additional Fisher information and exhibits constraining power similar to that of the \texttt{PS}. \textit{Right plot}: The \texttt{MPS}, in contrast, for mark \mref{m2}, successfully captures additional Fisher information with comparable constraining power to the \texttt{PS}. The cyan cross corresponds to fiducial values of the EoR model parameters.}
    \label{fig:Fisher_information_extract}
\end{figure}

\begin{itemize}

\item \textbf{Non-uniqueness and choice of mark}

There is no unique theoretical prescription for selecting an optimal mark. In practice, the choice is problem-dependent and often guided by heuristic considerations. A large functional freedom exists in defining marks, and even within a single functional form, smoothing scales and parameter choices effectively define different transformations. Optimality in the definition of mark is therefore not absolute but tied to the scientific objective and the specific metric used to evaluate its performance. Improvement in the constraints of parameter thus cannot be assumed a priori and must be demonstrated case by case. More importantly, marks such as \mref{m0} are not observables associated with the radio interferometric observations of the 21-cm signal since they require the mean of the signal. However, its simple functional form allows us to understand, interpret, and design additional robust functional forms such as \mref{m1} and \mref{m2} to increase the information content in the signal estimators. To add to this, apparent complex marks can provide better forecasts as compared to \texttt{PS}, as discussed in Appendix \ref{appendix:functions}.

\item \textbf{Degeneracies in transformation and mark parameter space}

While defining marks, mathematical degeneracies can arise within the transformation itself. Different combinations of mark parameters may produce apparently identical transformed maps. In such cases, the resulting marked power spectra are indistinguishable, even though the mark parameter values differ. Furthermore, distinct regions of the mark parameter space may yield nearly identical power spectra in both amplitude and shape. This introduces ambiguity in interpretation and complicates attempts to associate specific physical meanings with individual choices of parameters.

\item \textbf{Perturbation strength and stability concerns} 

If the transformation due to the mark is too aggressive, it may over-amplify nuisance components or suppress physically relevant structures. Excessive weighting can distort the 21-cm map's morphological features, potentially erasing valuable information rather than enhancing it, as seen in Figure \ref{fig:Combined_Power_spectrum_R_03_M3_compare_2.0.pdf}. Importantly, there is no objective stability criterion that defines a safe range of mark parameters. Determining when a transformation becomes overly destructive remains a practical challenge. 

Conversely, if the transformation is not significant enough, then the cost of applying additional transformations to the inference pipeline using the power spectrum becomes meaningless, or there may be a significant loss of information in the map itself. For this reason, we fixed the smoothing length scale at ${\rm R} = 1.68~{\rm Mpc}$, and did not explore smoothing scales larger than this to avoid loss of signal fluctuations at smaller (non-linear) scales.

\item \textbf{Mode mixing and interpretation}

Non-linear transformations and smoothing introduce non-trivial mode mixing. Real-space multiplication corresponds to convolution in Fourier space, redistributing power across $k$-modes. As a result, the Marked Power Spectrum cannot always be interpreted scale-by-scale in the same manner as the original spectrum. Additionally, different marked estimators may share significant overlapping information content. Simply combining multiple marked spectra may does not guarantee additive information gain and requires careful covariance modeling, to avoid adding redundant information.

\item \textbf{No guarantee of improvement}

Finally, marking does not automatically guarantee improved parameter constraints. Some stable and mathematically well-defined transformations may perform no better than the standard power spectrum. The 21-cm marked statistics may carry complementary information, but complementarity does not necessarily imply stronger constraints. The effectiveness of a mark depends critically on whether it emphasizes physically relevant structures for the parameters of interest, as observed in Figure \ref{fig:Fisher_information_extract}. 

\end{itemize}

Above mentioned points indicate that the 21-cm marked power spectrum has the potential to perform better than the standard power spectrum. Using suitable stable marks, one can characterize the statistical properties of the HI 21-cm differential brightness temperature intensity map. We demonstrate, for the first time, that the Marked Power Spectrum is sensitive to different environments of the IGM that are responsible for the fluctuations in the brightness temperature maps. Furthermore, the Fisher matrix analysis clearly shows that the 21-cm marked power spectrum carries additional non-Gaussian information than the standard power spectrum. Thus, the 21-cm marked power spectrum performs better than the standard power spectrum in describing the statistical properties of the IGM, capturing non-Gaussian information from the EoR HI 21-cm differential brightness temperature maps, and breaking model parameter degeneracies. 

In this work, we have not considered realistic systematics and detectability of the Marked power spectrum; for future work, we would like to study SKA-like instruments having the potential to probe 3D tomographic maps, including systematics and noise, and their impact on the detectability of the Marked power spectrum in more detail.

\acknowledgments
LN acknowledge the financial support by the Department of Science and Technology, Government of India, through the INSPIRE Fellowship [IF210392]. VP, LN and SM acknowledge the use of computing infrastructure for this work, which is hosted at the DAASE, IIT Indore and was procured through the funding via , Department of Science and Technology, Government of India sponsored DST-FIST grant no. SR/FST/PSII/2021/162 (C) awarded to the DAASE, IIT Indore. We would like to thank Manas Mohit Dosibhatla for useful discussions and suggesting the title of the paper.


\newpage

\appendix
\section{Evolution of 21-cm Marked Power Spectrum}
\label{appwndix:Evolution}

\begin{figure}[htbp]
\centering
\includegraphics[width=1\linewidth]{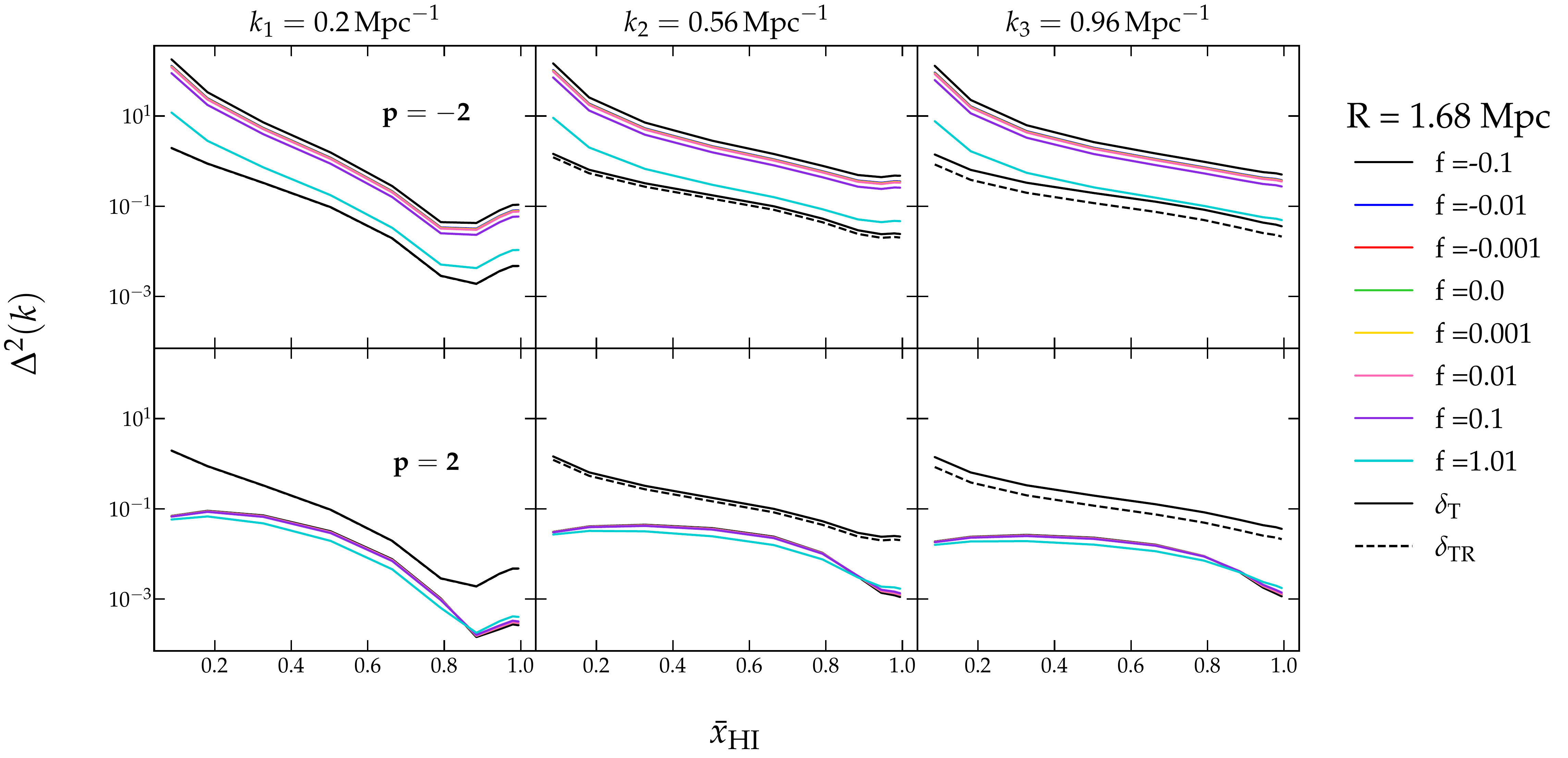}
\caption{Evolution of various \texttt{MPS} for \mref{m0} with mean neutral fraction. The solid black line is the power spectrum of the brightness temperature contrast, the dashed black line is the power spectrum for the corresponding smoothed field, and the other colored lines are the Mark Power spectra for a fixed functional form. We have chosen to compare the mark power spectra with the brightness temperature contrast to maintain consistent units.}
\label{fig:Power_evolution_R_03_p_evolution_delta_T_Mark_manual_m0fT.pdf}
\end{figure}

\begin{figure}[htbp]
\centering
\includegraphics[width=1\linewidth]{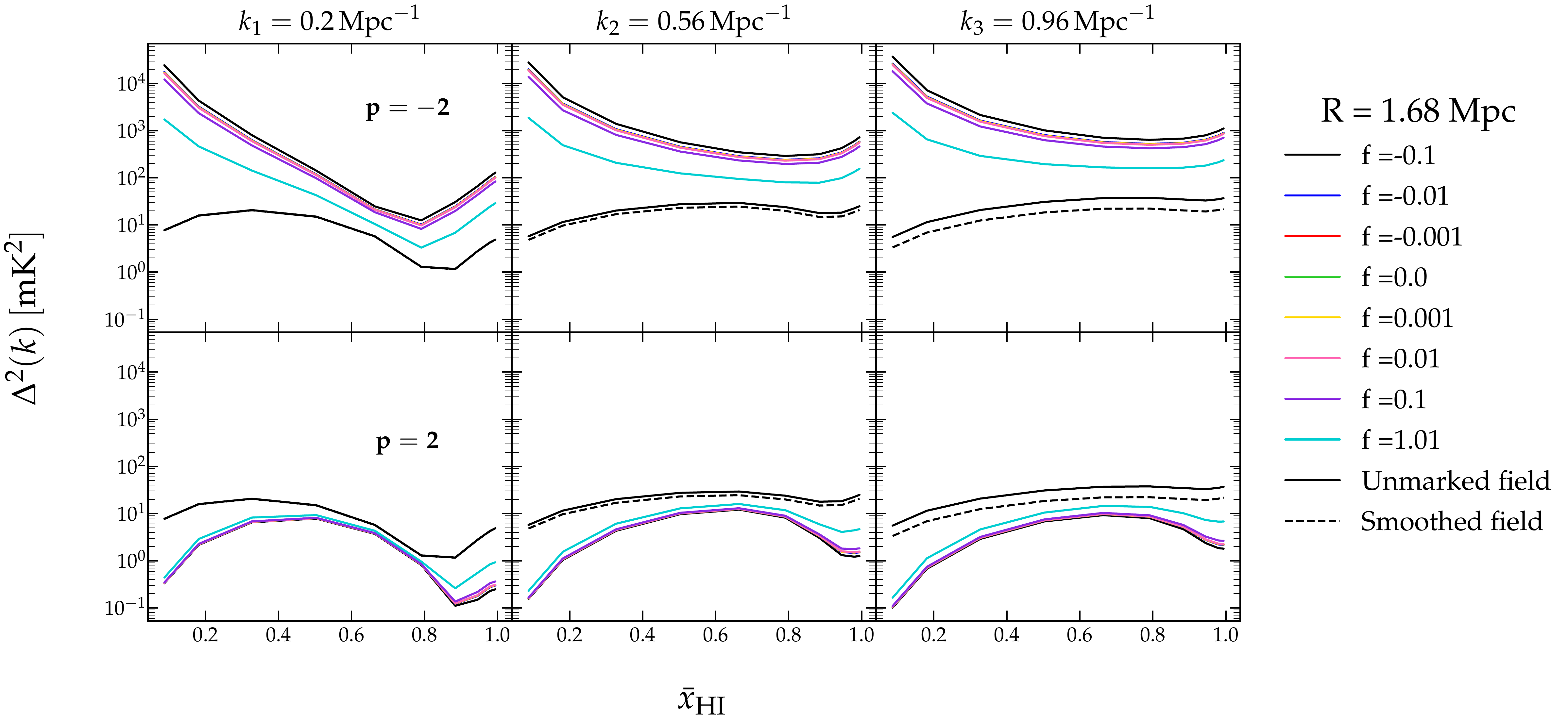}
\caption{Same as Figure \ref{fig:Power_evolution_R_03_p_evolution_delta_T_Mark_manual_m0fT.pdf} for \texttt{MFPS}. Here, we compare the power spectra of the mean subtracted brightness temperature maps with the marked field power spectrum to ensure consistent units comparison. }
\label{fig:Power_evolution_R_03_p_evolution_delta_T_Marked_manual_m0dT.pdf}
\end{figure}

In this section, we study the evolution of the 21-cm Marked Power Spectrum, along with the 21-cm power spectrum and the power spectra of the corresponding smoothed fields, for mark \mref{m0} with mass-averaged neutral fraction, for a fixed set of $k$-modes. We use the same simulation setup as discussed in previous sections and generate the HI 21-cm differential brightness temperature intensity maps for the following redshifts: $ z \in \{13.20, 11.78, 10.62, 9.65, 8.83, 8.13,\\ 6.9, 6.5, 6.1 \}$, including $z=7.52$, which has been used for the analysis of the work so far. Figures \ref{fig:Power_evolution_R_03_p_evolution_delta_T_Mark_manual_m0fT.pdf} and \ref{fig:Power_evolution_R_03_p_evolution_delta_T_Marked_manual_m0dT.pdf} show the evolution of \texttt{MPS} and \texttt{MFPS} respectively that are informed of ionized and neutral regions.

For a fixed value of the exponent $p$, we observe a clear trend in $f$, which is similar to that seen in Figures \ref{Combined_Power_spectrum_R_03_M3_compare_-2.0.pdf} and \ref{Combined_Power_spectrum_R_03_M5_compare_-2.0.pdf}, which is for other marks \mref{m1} and \mref{m2}. Interestingly, the trend flips when the sign of $p$ changes. This is also expected when comparing Figures \ref{Combined_Power_spectrum_R_03_M3_compare_-2.0.pdf} and \ref{fig:Combined_Power_spectrum_R_03_M3_compare_2.0.pdf}. This shows an important behavior in the choice of the mark as a general functional form, as well as the sign of the mark parameters. We observe a consistent behavior like the power spectrum: the amplitude and the shape change as explained in previous sections. However, we observe a loss of the characteristic curve of inside-out reionization in the evolution of the power spectra in Figure \ref{fig:Power_evolution_R_03_p_evolution_delta_T_Mark_manual_m0fT.pdf}, even for the unmarked maps. As one approaches recent redshifts, the mean brightness temperature decreases and thus from the form of \mref{m0}, dividing by this quantity yields higher power at recent redshifts, i.e., at lower mass-averaged neutral fraction. This becomes clear while looking at the power spectra evolution of the unmarked maps (solid black line) in Figure \ref{fig:Power_evolution_R_03_p_evolution_delta_T_Marked_manual_m0dT.pdf}, where we see the characteristic curve in the evolution of the power spectrum.

Moreover, when we assign this mark as a weight, we lose this curve feature for \texttt{MFPS} as well, just as for \texttt{MPS} for $p=-2$. Thus, even if \mref{m0}, as shown in the previous sections, is a very useful form, it apparently remains not so useful to study the evolution of the target signal. However, in the bottom panel of \ref{fig:Power_evolution_R_03_p_evolution_delta_T_Marked_manual_m0dT.pdf}, we also see the marked power spectra curve, as the fluctuations focus more on ionized regions ($p=2$), with less clustering and less power. This suppression of fluctuations helps when we reassign this mark to weights.

\begin{figure}[htbp]
\centering
\includegraphics[width=1\linewidth]{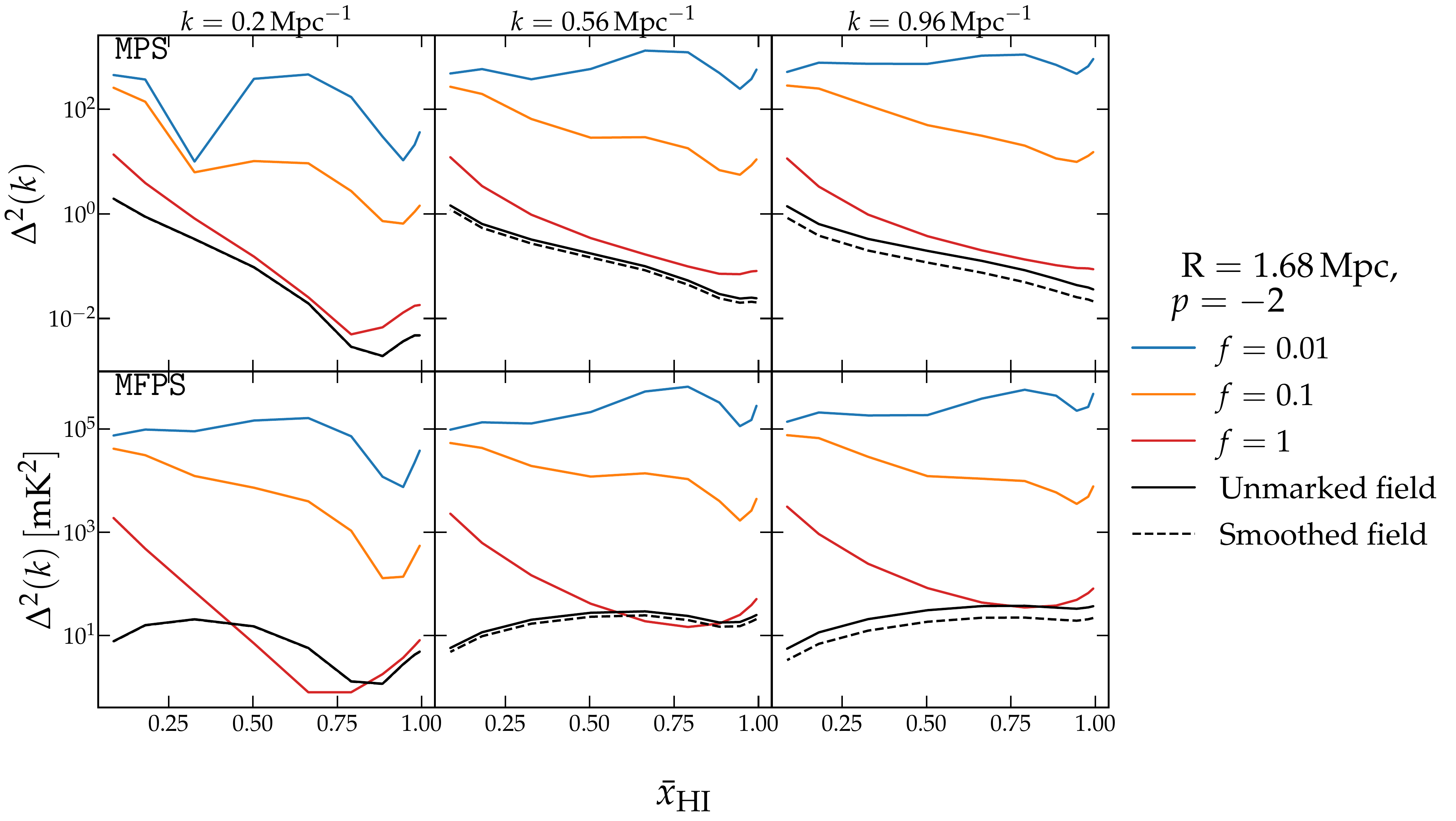}
\caption{Evolution of \texttt{MPS} (\textit{top panel}) and \texttt{MFPS} (\textit{bottom panel}), for mark \mref{m1}, for a representative cases of mark parameters values. The unmarked and smoothed maps in the top panel correspond to brightness temperature contrast $\delta_{{\rm T}_{\rm b}}$, and that for the bottom panel is the mean-subtracted brightness temperature map ${\rm dT}$.}  
\label{fig:Power_evolution_R_03_p_evolution_Marked_manual_m3.pdf}
\end{figure}

\begin{figure}[htbp]
\centering
\includegraphics[width=1\linewidth]{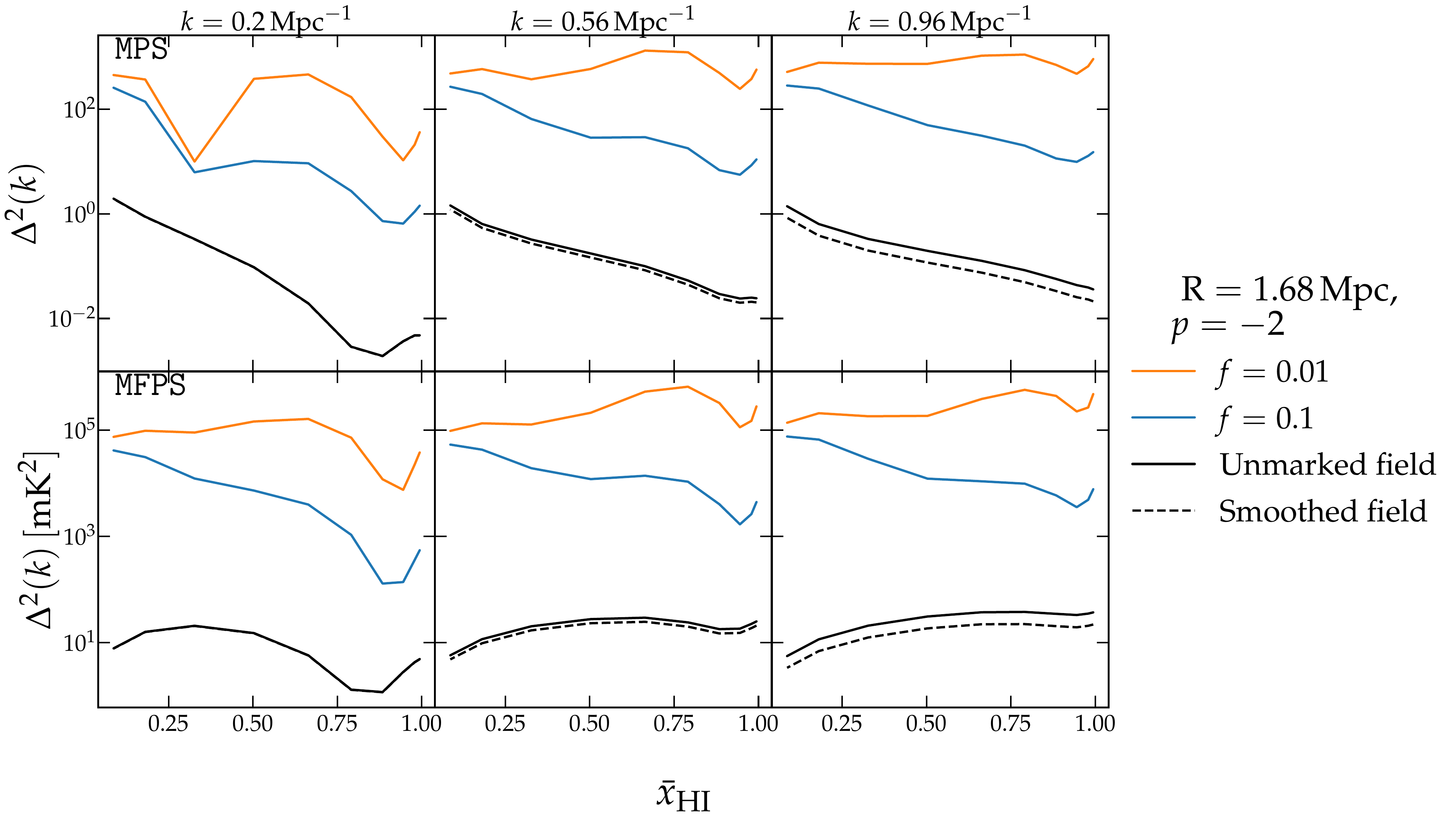}
\caption{Same as Figure \ref{fig:Power_evolution_R_03_p_evolution_Marked_manual_m3.pdf}, for mark \mref{m2}. }  
\label{fig:Power_evolution_R_03_p_evolution_Marked_manual_m5.pdf}
\end{figure}

Figures \ref{fig:Power_evolution_R_03_p_evolution_Marked_manual_m3.pdf} and \ref{fig:Power_evolution_R_03_p_evolution_Marked_manual_m5.pdf} show the evolution of the 21-cm marked power spectrum, for marks \mref{m1} and \mref{m2} respectively. We observe that for small $k$-modes, the characteristic inside-out reionization curve is evident in both chosen marks, particularly for \texttt{MPS}; however, it is not as sharp for larger $k$-modes. More importantly, we observe a clear degeneracy in the marked power spectra for these different marks, not for the different mark parameters of these. This again points out the motivation behind defining such marks, where, as mentioned in previous sections, the mark is constructed as a more general way to capture the non-Gaussian information content. With this motivation, this degeneracy between different functional forms is expected.  
Also, not all 21-cm marked power spectra capture the expected time evolution of the target signal, as seen in Figure \ref{fig:Power_evolution_R_03_p_evolution_Marked_manual_m3.pdf} for the red curve. It is unclear what this marked power spectrum is probing.

This again points to the heart of the study: the usefulness of a mark depends heavily on the problem statement at hand.

\section{Additional marks}
\label{appendix:functions}

In this section, we briefly discuss marks, other then the ones used in the analysis for far. Since finding the right mark lies at the heart of all the study it is worthwhile to have a look as the nature of these marks and their ability to capture the non-Gaussian information in the EoR 21-cm signal. 

\subsection{Functional forms}
\label{appendix:functional_forms}

As mentioned in previous sections, the functional form of the mark can vary depending on the problem statement and the nature of the map. We explore functional forms \mref{m1}, \mref{m2} as given. An important point to note here is that the functional form and hence, the corresponding power spectrum have units based on the values of mark parameters. \\

\begin{equation}\label{ma}
    m_a(\mathbf{x},z; \text{R},f,p) = \Bigg[ \frac{1 + f}{1 + f + \Big[ \text{dT}(\mathbf{x},z) \Big]_\text{R} }\Bigg]^p
\end{equation}

\begin{equation}\label{mb}
    m_b(\mathbf{x},z; \text{R},f,p) = \Bigg[ \frac{1 + f}{1 + f + \Big[ \delta\text{T}_\text{b}(\mathbf{x},z) \Big]_\text{R} }\Bigg]^p
\end{equation}

\begin{figure}[htbp]
    \centering

    \subcaptionbox{$m_a({\rm R}=1.68 \, {\rm Mpc}, f=0, p=2)$.\label{fig:ma_p2.png}}[0.45\linewidth]{
        \includegraphics[width=\linewidth]{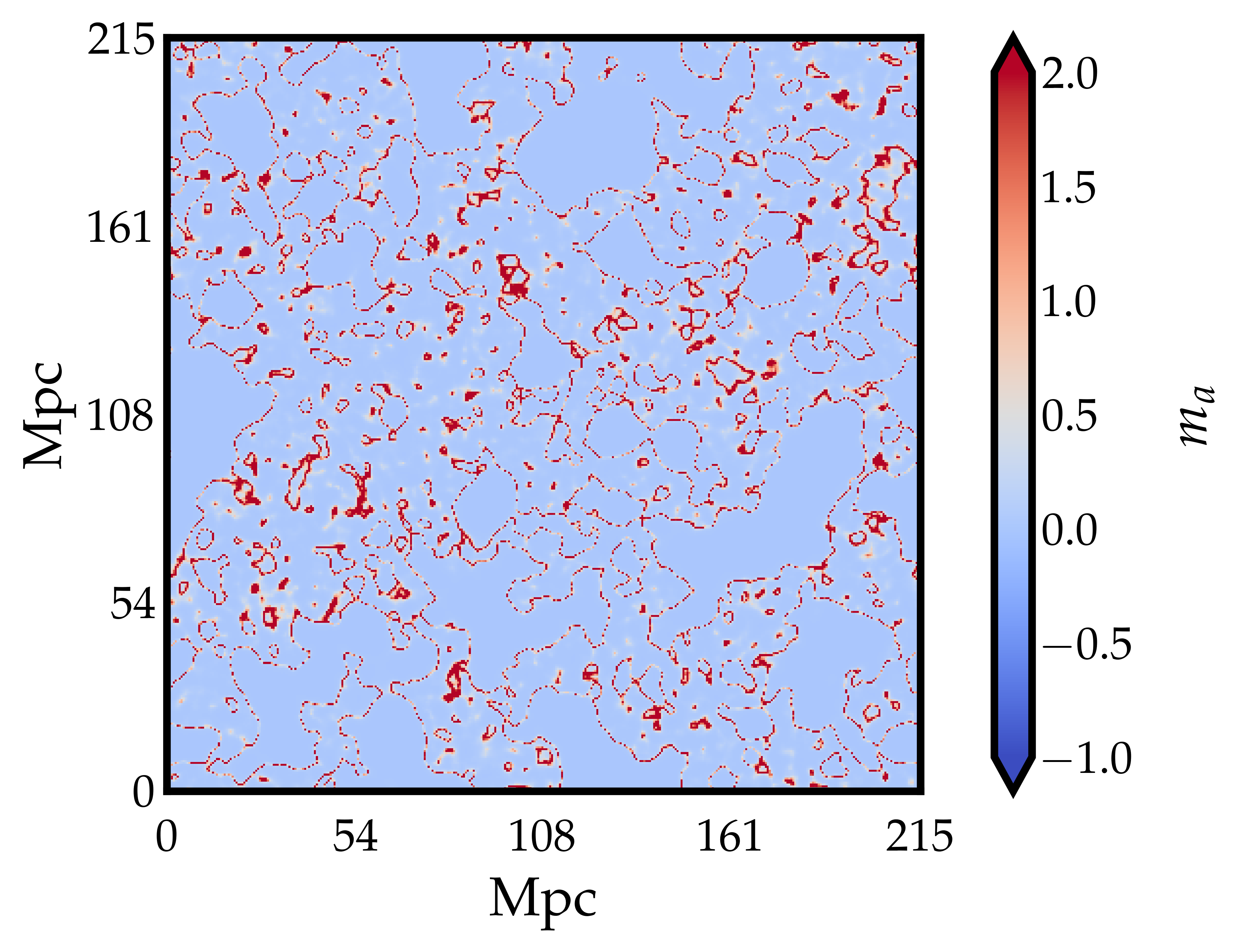}}
    \hfill
    \subcaptionbox{$m_a({\rm R}=1.68 \, {\rm Mpc}, f=0, p=-2)$.\label{fig:ma_p-2.png}}[0.45\linewidth]{
        \includegraphics[width=\linewidth]{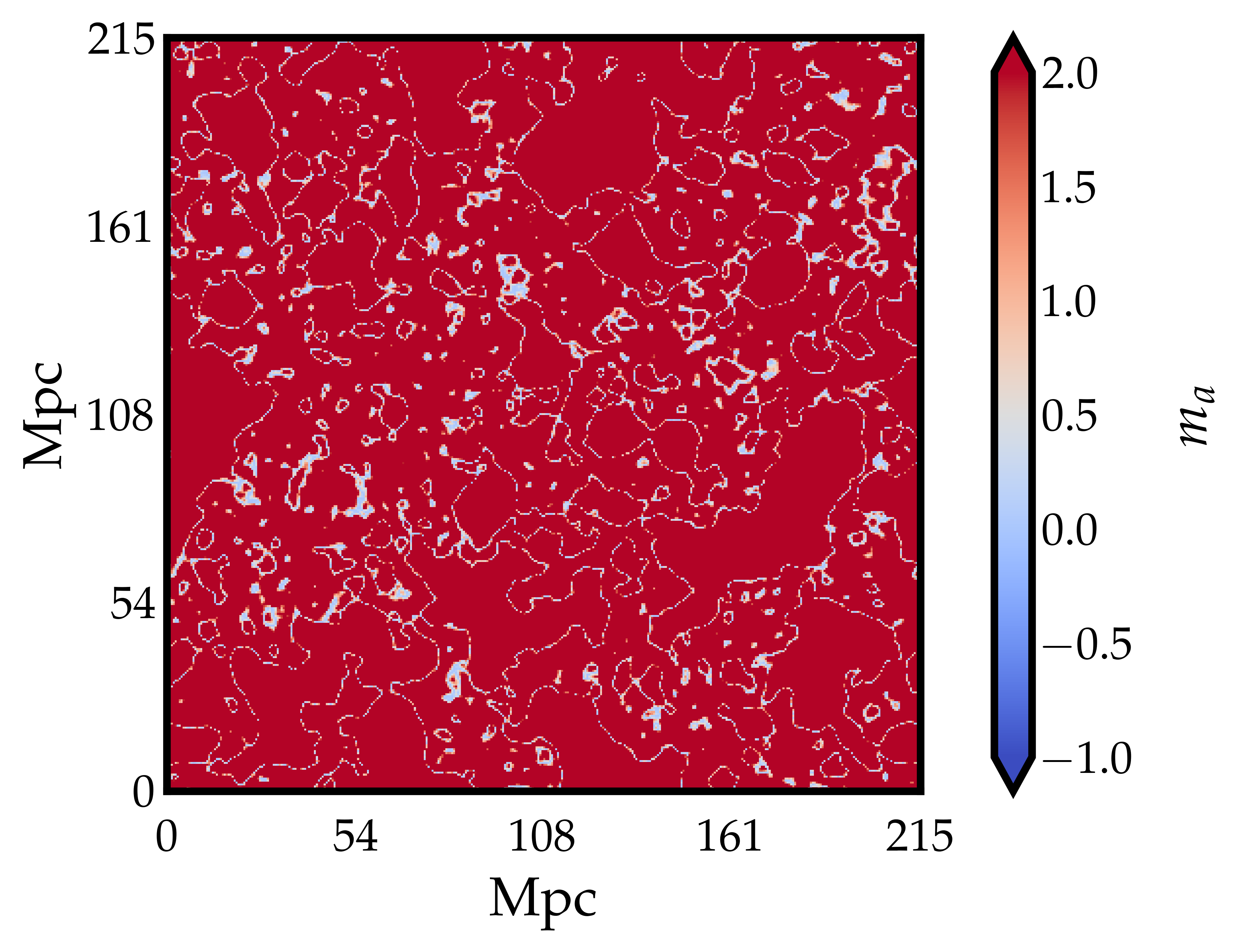}}
    
    \vspace{0.1cm}

    \subcaptionbox{$\mathbf{M}_a({\rm R}=1.68 \, {\rm Mpc}, f=0, p=2)$.\label{fig:maXdT_p2.png}}[0.45\linewidth]{
        \includegraphics[width=\linewidth]{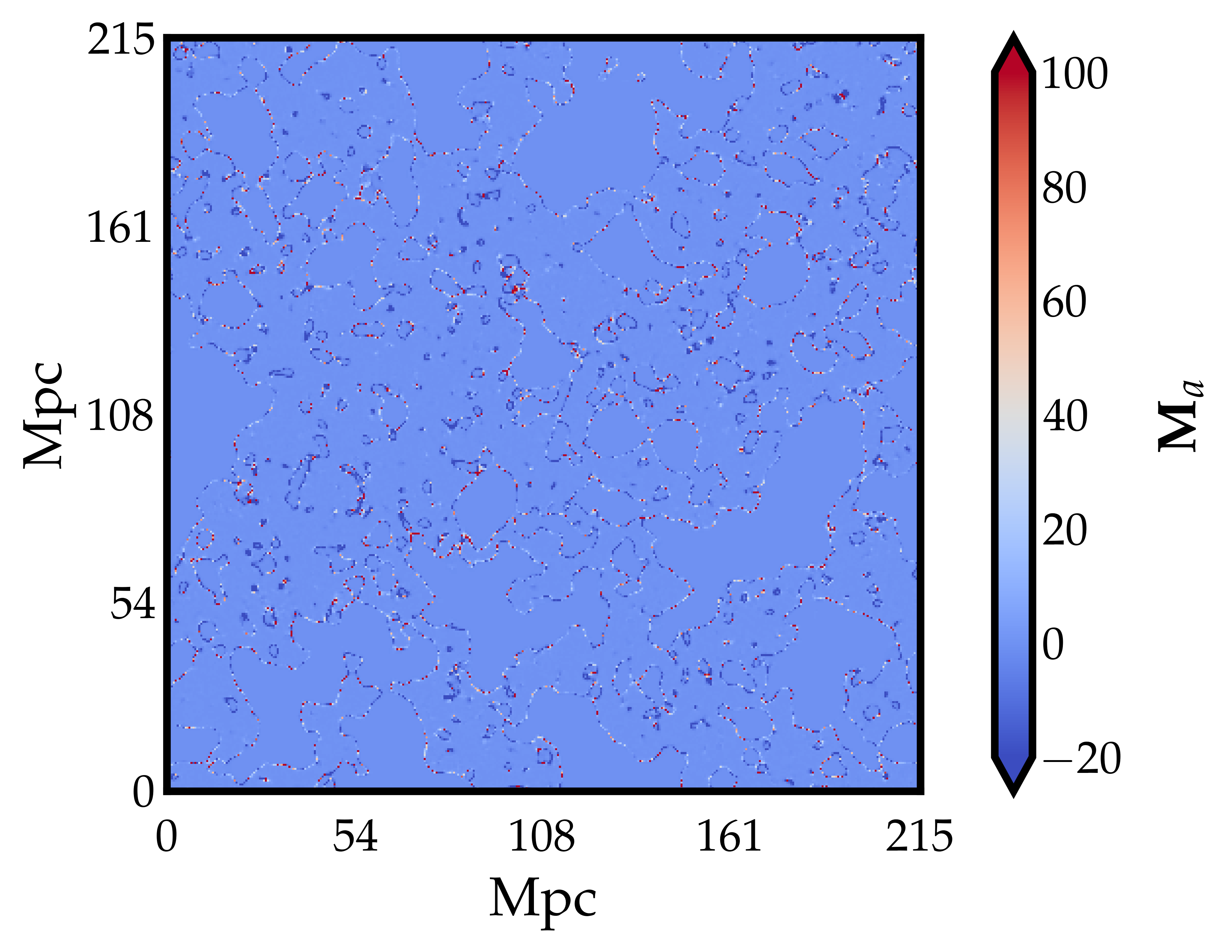}}
    \hfill
    \subcaptionbox{$\mathbf{M}_a({\rm R}=1.68 \, {\rm Mpc}, f=0, p=-2)$.\label{fig:maXdT_p-2.png}}[0.45\linewidth]{
        \includegraphics[width=\linewidth]{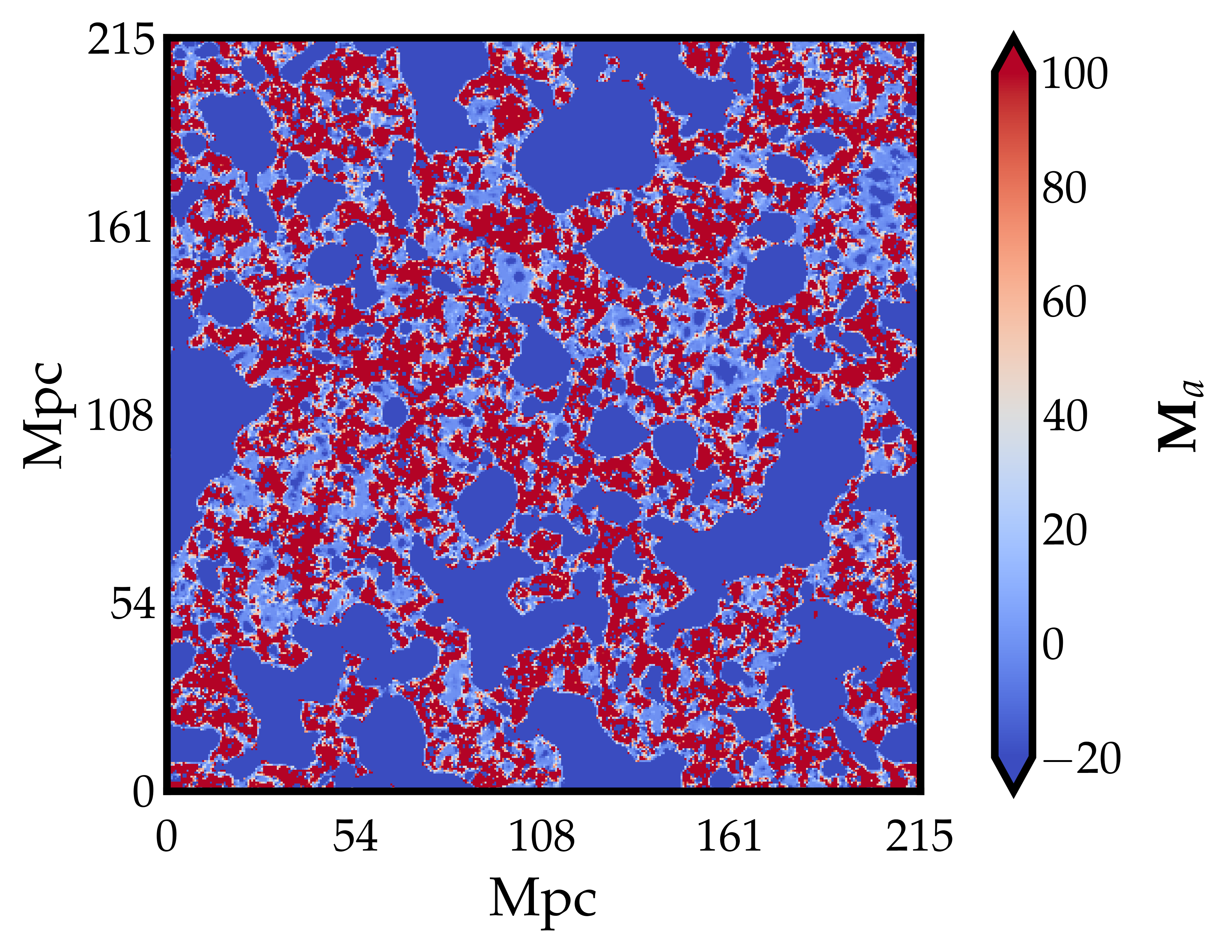}}
    
    \vspace{0.1cm}
    
    \subcaptionbox{$m_b({\rm R}=1.68 \, {\rm Mpc}, f=0, p=2)$.\label{fig:mb_p2.png}}[0.45\linewidth]{
        \includegraphics[width=\linewidth]{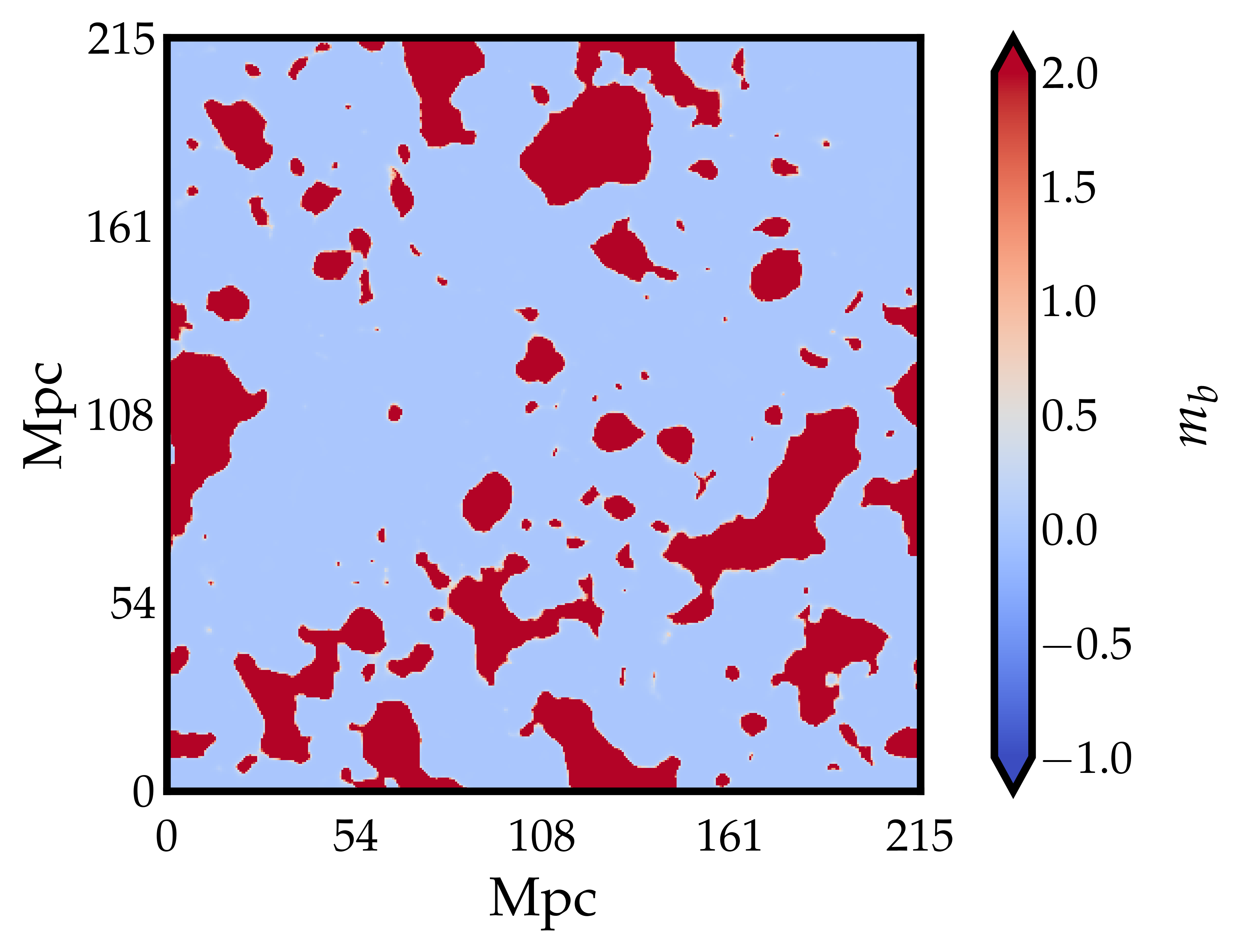}}
    \hfill
    \subcaptionbox{$m_b({\rm R}=1.68 \, {\rm Mpc}, f=0, p=-2)$.\label{fig:mb_p-2.png}}[0.45\linewidth]{
        \includegraphics[width=\linewidth]{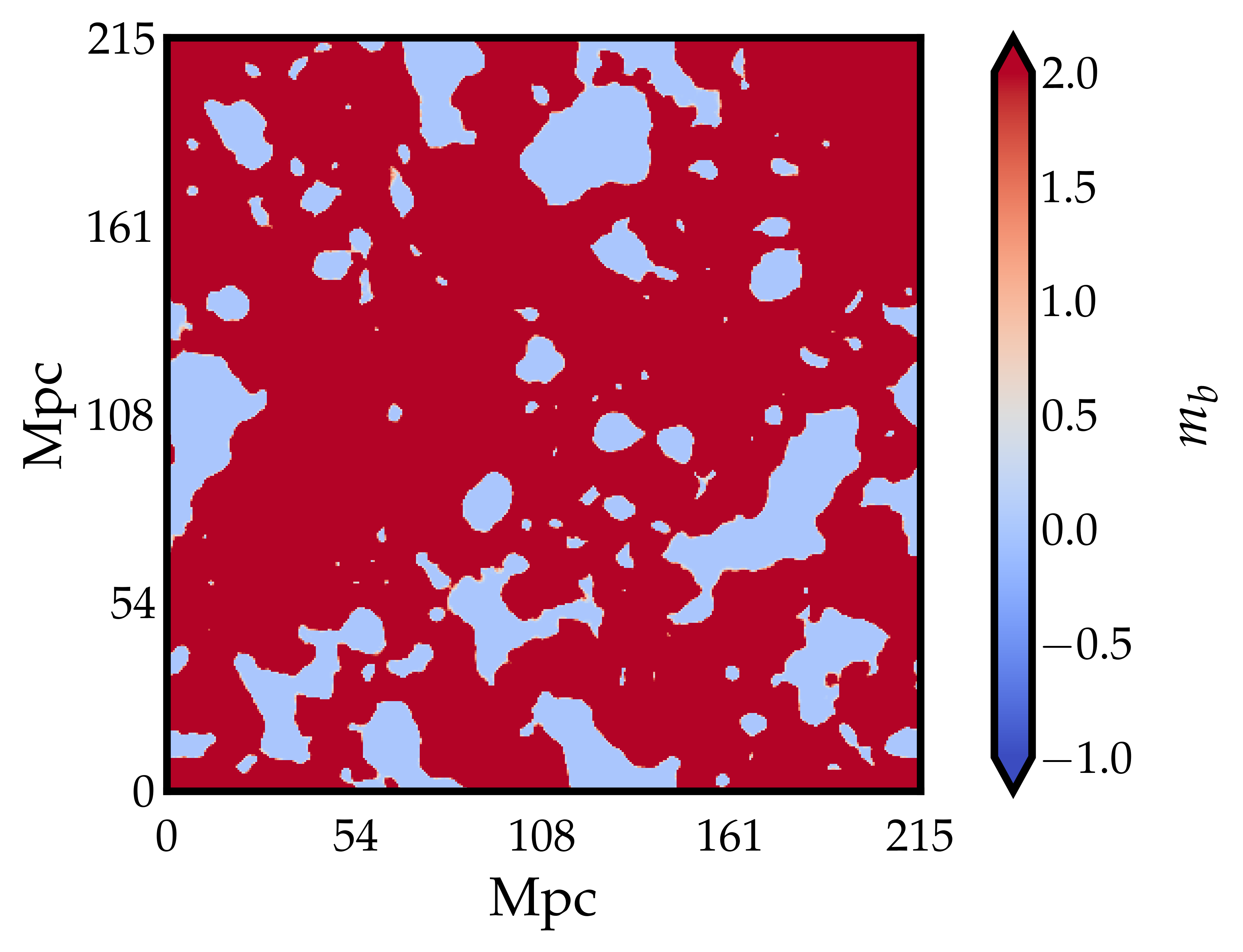}}
    
    \vspace{0.1cm}

    \subcaptionbox{$\mathbf{M}_b({\rm R}=1.68 \, {\rm Mpc}, f=0, p=2)$.\label{fig:mbXdT_p2.png}}[0.45\linewidth]{
        \includegraphics[width=\linewidth]{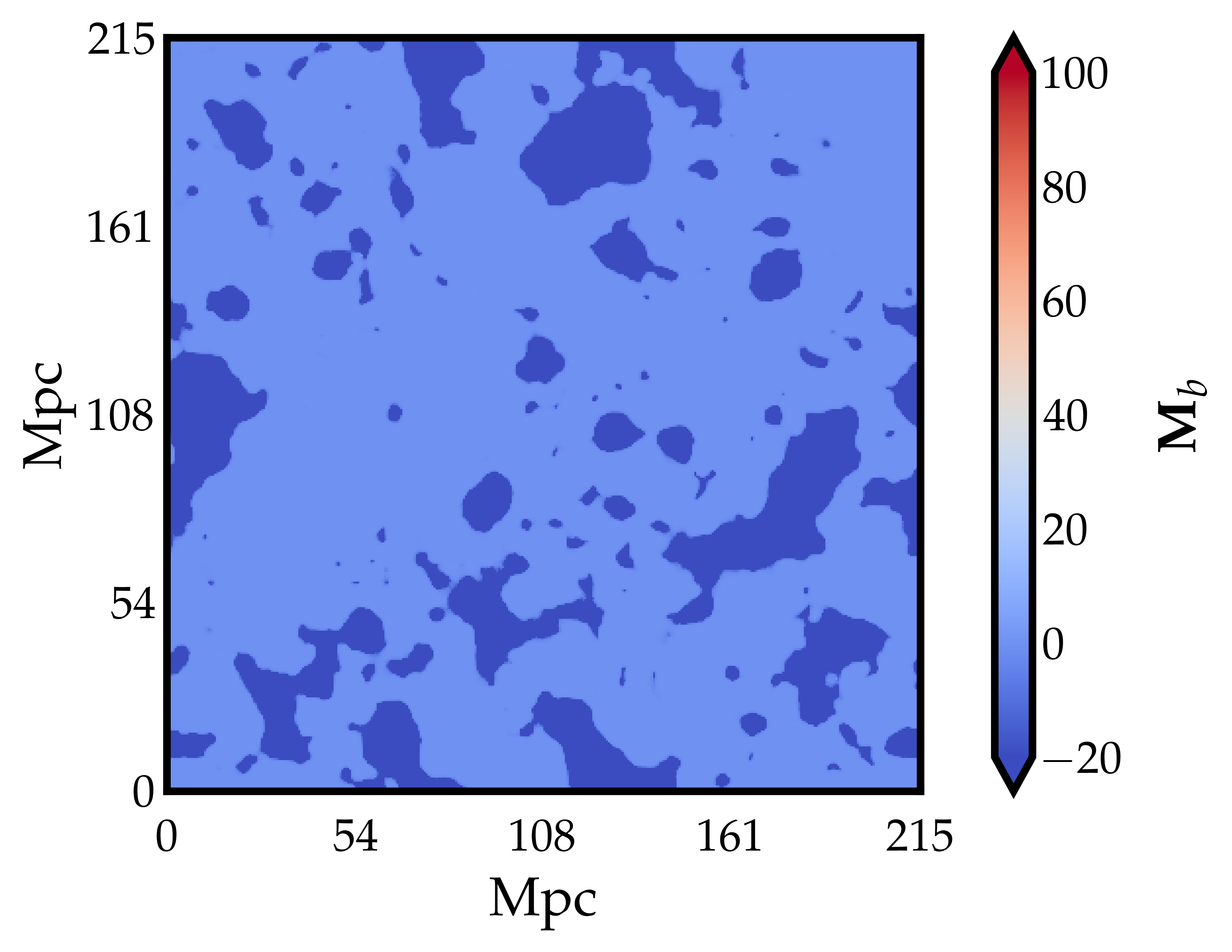}}
    \hfill
    \subcaptionbox{$\mathbf{M}_b({\rm R}=1.68 \, {\rm Mpc}, f=0, p=-2)$.\label{fig:mbXdT_p-2.png}}[0.45\linewidth]{
        \includegraphics[width=\linewidth]{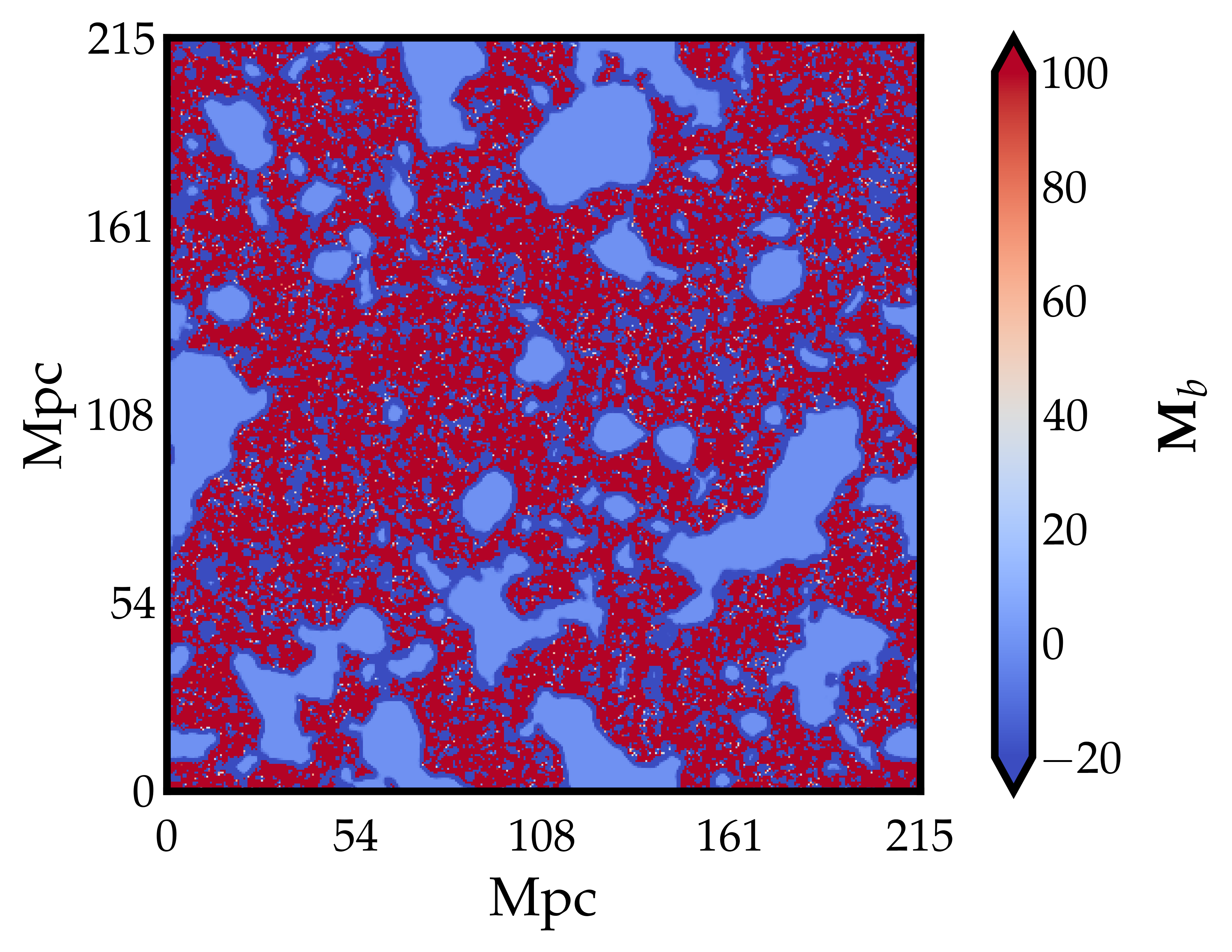}}

    \caption{Same as Figure \ref{fig:combined_brightness_temperature_marks} for marks $m_a$ and $m_b$. The units of these quantities depend on the choice of free parameters unlike previously defined marks.}
    
    \label{fig:mamb_2D_plots}
\end{figure}

Figure \ref{fig:mamb_2D_plots} shows 2D slices of marks $m_a$ and $m_b$, at $z=7.52$, for representative cases of mark parameters. As seen from the figure, these marks respond very differently to the input intensity map. Transformations such as \ref{fig:ma_p2.png}, \ref{fig:ma_p-2.png}, and \ref{fig:maXdT_p2.png} do not seem very useful, as they perturb the input map to such an extent that the loss of distinct environmental features becomes evident. Complementary transformations, such as \ref{fig:mb_p2.png} and \ref{fig:mb_p-2.png}, visually flip the contrast between distinct environments and thus have the potential to tune and capture their statistical properties. Transformations such as \ref{fig:maXdT_p-2.png}, \ref{fig:mbXdT_p2.png}, and \ref{fig:mbXdT_p-2.png} preserve the broader signal structure while enhancing or suppressing features in the IGM. More importantly, unlike the stable marks defined in earlier sections, these marks are not unitless. Instead, the unit of these marks (and the marked field) depends on the choice of mark parameters. This poses a major challenge for interpreting results and drawing conclusions when using these marks in the analysis, which is why we restrict our analysis to specific marks to ensure a selective, stable choice of free parameters.

\subsection{Fisher matrix analysis}
\label{appendix:Fisher}

In this section, we briefly discuss the Fisher forecast in the EoR model parameters using \texttt{MPS} for marks $m_a$ and $m_b$. 

\begin{figure}[htbp]
\centering

\begin{minipage}{0.48\linewidth}
\centering
\includegraphics[width=\linewidth]{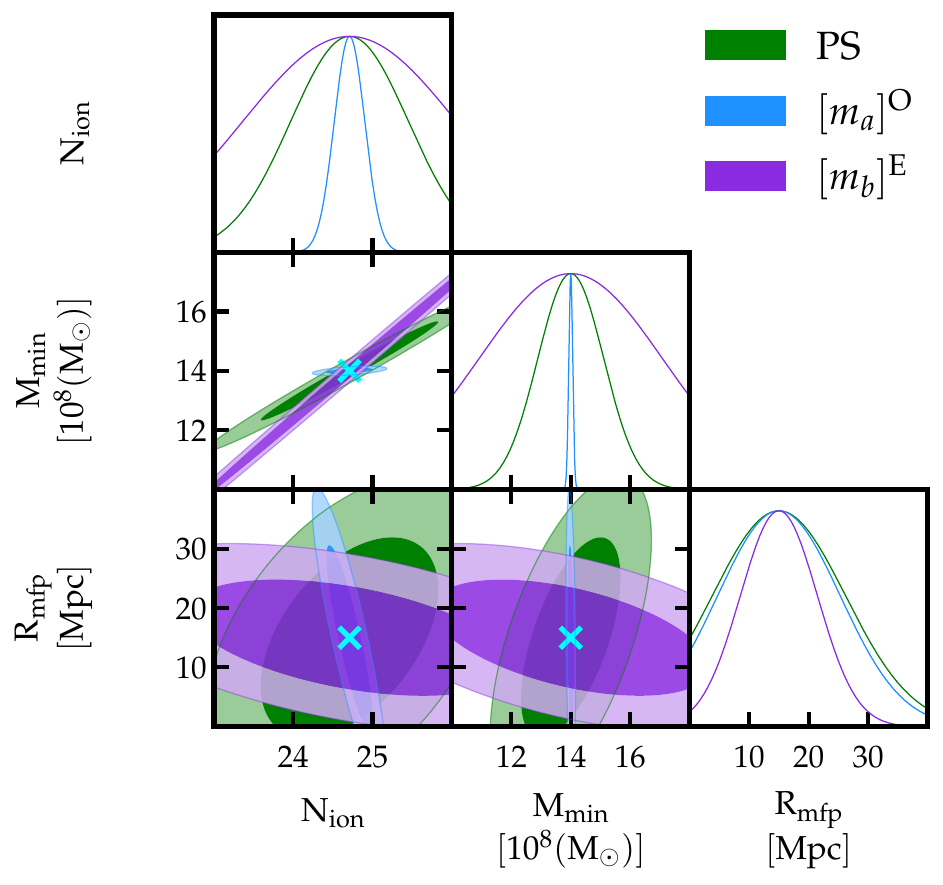}
\end{minipage}
\hfill
\begin{minipage}{0.48\linewidth}
\centering
\includegraphics[width=\linewidth]{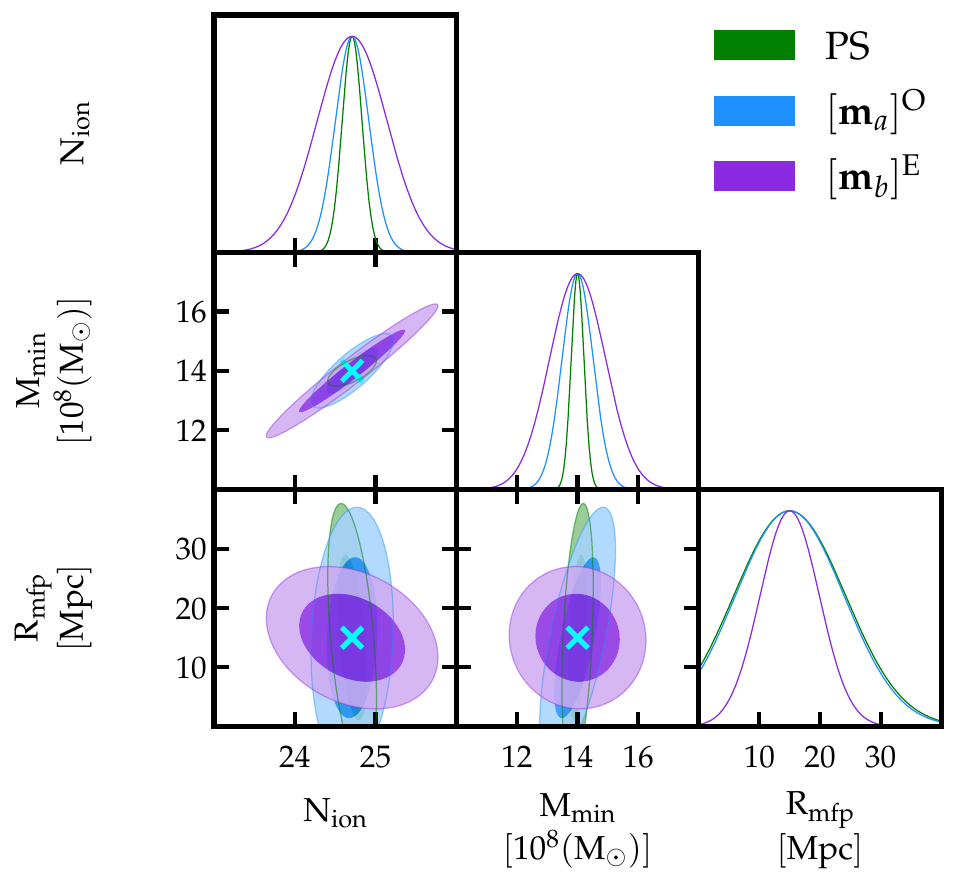}
\end{minipage}

\caption{Fisher forecast of EoR model parameters using marks $m_a$ and $m_b$ (\textit{Left plot}) and their normalized versions (\textit{right plot}). The mark parameter values chosen that gives stable power spectra are as given: $m_a({\rm R}=1.68 , {\rm Mpc}, f=0, p=-2)$ and $m_b({\rm R}=1.68 , {\rm Mpc}, f=0.001, p=2)$}
\label{fig:Fisher_m1_m2}
\end{figure}

The left plot of Figure \ref{fig:Fisher_m1_m2} compares the Fisher forecast using \texttt{PS} and \texttt{MPS} for these marks, where the choice of mark parameters gives stable power spectra. We clearly observe smaller errors in the forecast of model parameters using $m_a$ and successful capture of complementary information using $m_b$. The drawback of using such marks is that the units are arbitrary: the unit of the mark (and hence the corresponding marked power spectrum) depends on the choice of mark parameter. This creates ambiguity in interpreting the result. One can ignore the units and use the marked power spectra as a statistical tool to capture the information content of the signal, without interpreting why they provide a better edge than previously defined marks.

In the simplest case, one can use such marks, normalize them by their mean, and then use the Marked Power Spectrum, as shown in the right plot of Figure \ref{fig:Fisher_m1_m2}. Hence, $\mathbf{m}_j(\mathbf{x}, z : {\rm R}, f, p) = m_j(\mathbf{x}, z : {\rm R}, f, p)/\bar{m}_j(z : {\rm R}, f, p)$ where $j=a,b$ and $\bar{m}_j$ is the mean of $m_j$ over the simulation volume ${\rm V}$. One can use such normalizations or other normalizations, such as the mark used in \cite{cowell2025constraintsmarkedangularpower}, where a linear mark is normalized by the standard deviation and reweighted to the maps.

In the context of the 21-cm marked power spectrum, one expects such marks, after normalization by their mean, to behave similarly to the un-normalized version, which is the case to some degree as seen in the figure. However, we observe a significant deterioration in the model parameter forecasts using these normalized versions. One explanation is the loss of dynamic range in the power spectra, which reduces sensitivity to EoR model parameters with strong power spectra. However, we observe that these statistics capture complementary information from the map and can break model parameter degeneracy.








\bibliographystyle{JHEP}
\bibliography{biblio}
\end{document}